\documentclass[a4paper,12pt]{article}
\usepackage{authblk}
\usepackage[square,sort,comma,numbers]{natbib}
\usepackage{soul}
\usepackage{color}
\usepackage{pifont}
\usepackage{mathptmx}
\usepackage{graphicx}
\usepackage{amsmath, amsfonts, amssymb, mathrsfs}
\usepackage{caption}
\usepackage{multirow}
\usepackage[normalem]{ulem}
\useunder{\uline}{\ul}{}
\usepackage{longtable}
\usepackage{tabu}
\usepackage{fixmath}

\DeclareMathOperator*{\argmax}{arg\,max}

\title{Performance and Application of Estimators for the Value of an Optimal Dynamic Treatment Rule}
\author[1]{Lina Montoya \thanks{Email address for correspondence: lmontoya@unc.edu}}
\affil[1]{University of North Carolina, Chapel Hill, Department of Biostatistics}
\author[2]{Jennifer Skeem}
\affil[2]{University of California, Berkeley, Departments of Social Welfare and Public Policy}
\author[3]{Mark van der Laan}
\affil[3]{University of California, Berkeley, Division of Biostatistics}
\author[3]{Maya Petersen}
\date{October 2021}

\begin{document}

\maketitle
\pagebreak

\begin{abstract}

Given an (optimal) dynamic treatment rule, it may be of interest to evaluate that rule -- that is, to ask the causal question: what is the expected outcome had every subject received treatment according to that rule? In this paper, we study the performance of estimators that approximate the true value of: 1) an \textit{a priori} known dynamic treatment rule 2) the true, unknown optimal dynamic treatment rule (ODTR); 3) an estimated ODTR, a so-called “data-adaptive parameter," whose true value depends on the sample. Using simulations of point-treatment data, we specifically investigate: 1) the impact of increasingly data-adaptive estimation of nuisance parameters and/or of the ODTR on performance; 2) the potential for improved efficiency and bias reduction through the use of semiparametric efficient estimators; and, 3) the importance of sample splitting based on CV-TMLE for accurate inference. In the simulations considered, there was very little cost and many benefits to using the cross-validated targeted maximum likelihood estimator (CV-TMLE) to estimate the value of the true and estimated ODTR; importantly, and in contrast to non cross-validated estimators, the performance of CV-TMLE was maintained even when highly data-adaptive algorithms were used to estimate both nuisance parameters and the ODTR. In addition, we apply these estimators for the value of the rule to the ``Interventions" Study, an ongoing randomized controlled trial, to identify whether assigning cognitive behavioral therapy (CBT) to criminal justice-involved adults with mental illness using an ODTR significantly reduces the probability of recidivism, compared to assigning CBT in a non-individualized way.

\end{abstract}

\pagebreak

\section{Introduction}

There is an interest across disciplines in using both experimental and observational data to uncover treatment effect heterogeneity and quantifying the benefits of responding to this heterogeneity when assigning treatments (for example, \cite{khoury2016precision, laber2017dynamic}). Various methods aimed at estimating heterogenous treatment effects (HTEs) aim to answer the question, “who benefits from which treatment?” One way to uncover HTEs is by using the dynamic treatment rule framework. A dynamic treatment rule is any rule that assigns treatment based on covariates \citep{bembomvdL2007, vdLpetersen2007, robins1986new, chakraborty2013, Chakraborty2014}. An optimal dynamic treatment rule (ODTR) is the dynamic treatment rule that yields the highest expected outcome (if higher outcomes are better) \citep{murphy2003, robins2004, moodie2007}. In recent years, there has been a increase in literature describing methods to estimate the ODTR, from regression-based techniques to direct-search techniques; see, for example, \cite{kosorok2019}, \cite{kosorok2015adaptive}, and \cite{tsiatis2019dynamic} for recent overviews of the ODTR literature. One example of a data-adaptive method for estimating the ODTR is the SuperLearner algorithm, an ensemble machine learning approach that aims to best combine a library of candidate treatment rule estimators to work in tandem to yield the ODTR \citep{van2007super, luedtkeSLODTR, coyle2017computational}. In a companion paper, we review this approach, and highlight and investigate the consequences of key choices when implementing this method \citep{montoyaplaceholder}.

Once one knows or estimates a rule, it may be of interest to \emph{evaluate} it, which translates to asking the causal question: what is the expected outcome had every person received the treatment assigned to him or her by the (optimal) rule? The causal parameter that answers this question is sometimes referred to as the \emph{value} of the rule. It may be of relevance to learn this quantity in order to determine the benefit of assigning treatment in a more complex way compared to, for example, simply giving everyone treatment (an intervention that is straightforward to implement without the cost or complexity of measuring covariates and personalizing treatment assignment).

In this paper, we examine the following causal parameters, which we identify as statistical estimands, corresponding to the value of an (optimal) rule: 1) the true expected outcome of a given \textit{a priori} known dynamic treatment rule; 2) the true expected outcome under the true, unknown ODTR -- a particularly challenging target parameter to estimate; and 3) the true expected outcome under the \emph{estimated} ODTR, a so-called ``data-adaptive parameter." The latter parameter can be further split into the true expected outcome under a) an ODTR estimated on the entire data at hand, or b) a sample-split-specific ODTR, in which, under a cross-validation scheme, the ODTR is estimated on each training set and evaluated, under the true data-generating distribution, on the complementary validation set, with the data-adaptive parameter defined as an average across sample splits.

We discuss several estimators for these estimands. Specifically, we consider the following estimators suited for estimating a treatment-specific mean: the simple substitution estimator of the G-computation formula \citep{robins1986new}, the inverse probability of treatment weighted (IPTW) estimator \citep{hernanrobins2006, rosenbaum1983central}, the double-robust IPTW estimator (IPTW-DR) \citep{robins1994estimation, scharfstein1999theory, robins2000robust}, the targeted maximum likelihood estimator (TMLE) \citep{bembomvdL2007, rosenblum2010targeted, TLBBD, van2015targeted}, and the cross-validated TMLE (CV-TMLE) \citep{zheng2010asymptotic, van2015targeted, van2018targeted}. %Briefly, the CV-TMLE obtains initial fits of the estimator on the training set, updates the fit on the validation set (as in the TMLE procedure), and averages these updated validation set fits to generate an estimate of the parameter of interest. 

First, we review the conditions under which asymptotic linearity is achieved for these estimators in the scenario where one wants to evaluate an \textit{a priori} known rule. This provides insight into the common scenario in which one wishes to evaluate the value of a dynamic treatment rule that is pre-specified (based on investigator knowledge or external data sources), rather than learned from the data at hand. Estimators for this parameter require fast enough convergence rates and smoothness assumptions on nuisance parameters, though smoothness assumptions can be relaxed when employing CV-TMLE.

Second, we examine the more ambitious goal of estimating the expected outcome under the true, unknown ODTR, which additionally requires fast enough convergence of the estimate of the ODTR to the true ODTR, and for non cross-validated estimators, smoothness assumptions on ODTR estimators. Obtaining inference for the mean outcome under the ODTR has been shown to be difficult due to its lack of smoothness \citep{chakraborty2013, robins2004, laberchapter2017}; however, several methods have been proposed for constructing valid confidence intervals for this parameter, such as re-sampling techniques \citep{chakraborty2010, sies2019, chakraborty2013}. One approach to inference is to rely on algorithms based on parametric models; however, misspecification of these models can bias results. CV-TMLE relaxes the smoothness assumptions needed for inference, allowing one to use a single data set to safely estimate relevant parts of the data distribution (e.g., estimate nuisance parameters and/or the ODTR) and retain valid inference for the target parameter itself (e.g., the mean outcome under the ODTR) \citep{van2015targeted}. Such internal sample splitting is particularly important if the nuisance parameters or ODTR depend on a high dimensional covariate set or make use of data-adaptive methods \citep{van2015targeted}. 

Finally, it may instead be of interest to estimate the true outcome under an estimated ODTR (a data-adaptive parameter) because, in practice, it is the estimated rule that will likely be employed in the population, not the true rule, which is likely unknown \citep{van2015targeted, hubbard2016}. This relaxes the need for the estimate of the ODTR to converge to the true rule at a fast enough rate. Non-cross-validated estimators of this data-adaptive parameter still require smoothness assumptions on the estimate of the ODTR (and nuisance parameters) for asymptotic linearity. The use of CV-TMLE eliminates these requirements. This means that, at the cost of targeting a distinct, sample-split-specific target parameter, in a randomized experiment, achievement of asymptotic linearity for CV-TMLE with respect to the sample-split-specific data-adaptive parameter only requires that the estimated ODTR converges to a fixed rule \citep{van2015targeted}.

Previous simulation experiments have studied the performance of different estimators for the aforementioned statistical estimands in the setting in which a binary treatment is randomly assigned at a single time point. \cite{van2015targeted} demonstrated the importance of using an estimator of the value of the rule that uses a targeted bias reduction, such as TMLE and CV-TMLE, in order to improve performance. Of note, when evaluating the estimated rule, the %initial estimate of for the nuisance parameters needed for estimating the value of the rule, 
authors used the true treatment mechanism and, as an initial estimate of the outcome regression, either the true outcome regression or a constant value (i.e., an incorrectly specified outcome regression) when employing the (CV-)TMLE. \cite{coyle2017computational} extended these results by ``fully" estimating the value of the optimal rule, meaning the nuisance parameters were additionally estimated for both the optimal rule and the value of the rule, using the ensemble machine learning approach SuperLearner \citep{van2007super}. Both \cite{van2015targeted} and \cite{coyle2017computational} found that, indeed, there exists a positive finite sample bias when using TMLE versus CV-TMLE when estimating the value of the ODTR; in other words, with the rule learned and evaluated on the same data, estimates of the value of the rule may be optimistic, and CV-TMLE corrects this bias. Additionally, recently, \cite{sies2019} showed that cross-validation techniques for estimating the value of the rule, and in particular CV-TMLE, yielded a smaller difference between the true expected value under the true rule and its estimate, versus, for example, bootstrap techniques for evaluating a rule.

The current paper builds on previous work by illustrating, through a simulation study, how the degree of overfitting when estimating the optimal rule and/or nuisance parameters affects the performance of the estimators used for evaluating a rule. We also explore the potential for efficiency improvement and bias reduction through the use of semiparametric efficient estimators, with and without targeting. Finally, we show the importance of sample splitting using CV-TMLE when estimating the aforementioned statistical parameters.

We apply these estimators of the value of the rule to the Correctional Intervention for People with Mental Illness, or ``Interventions," trial, a ongoing study in which criminal justice-involved adults with mental illness -- a heterogeneous group with diverse symptoms, risk factors, and other treatment-relevant characteristics \citep{skeem2011correctional, skeem2014offenders} -- are either randomized to cognitive behavioral therapy (CBT) or treatment as usual (TAU).  Re-arrest, the outcome, is collected one year after randomization occurs, as a measure of recidivism. In our companion paper, we estimated the ODTR using the ODTR SuperLearner algorithm \citep{montoyaplaceholder} to identify which patients should receive CBT versus TAU. In this paper, we use CV-TMLE to determine whether administering CBT using the estimated ODTR is more effective in reducing recidivism than assigning CBT in a non-individualized way (for example, giving CBT to all offenders). 

%Specifically, we refer the reader to that companion paper for a description of the ODTR and show methods for estimating it -- in particular, using the SuperLearner ODTR -- and practical considerations and performance results for different configurations of that algorithm. In addition, as previously mentioned, in our companion paper we illustrate how to implement the SuperLearner using the ``Interventions" Study as an example. 

This article steps through the roadmap for answering causal questions \citep{petersen2014causal}, and is organized as follows. In the first section, we define the data and causal model, define the causal parameters as functions of counterfactual distributions contained in the causal model, and identify the statistical estimands as functions of the observed data distribution. In section 2 we discuss estimation, and in section 3 we discuss inference procedures and conditions for asymptotic linearity. In section 4 we present a simulation study illustrating the performance of these estimators. In section 5 we evaluate the ODTR SuperLearner algorithm that was applied to the ``Interventions" Study. Finally, we close with a discussion and future directions. In the Appendix, we provide a Notation Table with terms frequently used throughout this manuscript and our companion manuscript, in addition to definitions of each of the terms, as a reference to reader.

\section{Causal Roadmap}\label{sec2}

In this section, we follow the first steps of the roadmap for answering the causal questions: what would have been the expected outcome had everyone been given treatment according to: 1) any given rule; 2) the true ODTR; and 3) an estimate of the ODTR, which could either be a) a sample-specific estimate of the ODTR (i.e., an ODTR estimated on the entire sample), or b) a sample-split-specific estimate of the ODTR?

\subsection{Data and Models}

Structural causal models (SCM, denoted $\mathcal{M}^F$) will be used to describe the process that gives rise to variables that are observed (endogenous) and not observed (exogenous). The random variables in the SCM follow the joint distribution $P_{U,X}$; the SCM describes the set of possible distributions for $P_{U,X}$. The endogenous variables are the covariates $W \in \mathcal{W}$, binary treatment $A \in \mathcal{A} = \{0,1\}$, and outcome $Y \in \mathbb{R}$.  Exogenous variables are denoted $U = (U_W, U_A, U_Y)$.  The following structural equations illustrate dependency between the variables:
\begin{align*}
    W &= f_W(U_W), \\
    A &= f_A(U_A, A), \\
    Y &= f_Y(U_Y, A, W).
\end{align*}
Because we will be focusing on data where treatment is randomly assigned (as in the ``Interventions" trial), the above model can be modified by letting $U_A \sim Bernoulli(p=0.5)$ and $A=U_A$.   

We assume the observed data $O_i \equiv (W_i, A_i, Y_i) \sim P_0 \in \mathcal{M}$, $i = 1, \ldots, n$ were generated by sampling $n$ independent and identically distributed (i.i.d.) times from a data-generating system contained in the SCM $\mathcal{M}^F$ above. Here, $P_0$ is the observed data distribution, an element of $\mathcal{M}$, the statistical model.

The density of $O$ can be factorized as $p_0(O) = p_{W,0}(W) g_0(A|W) p_{Y,0}(Y|A,W),$ where $p_{W,0}$ is the true density of $W$, $g_0(A|W)$ is the true conditional probability of the treatment $A$, and $p_{Y,0}$ is the true conditional density of $Y$.

The empirical distribution $P_n$ gives each observation weight $\frac{1}{n}$. Estimates from this empirical distribution are denoted with a subscript $n$. If $V$-fold cross-validation is employed, the empirical data are uniformly and at random split into $V$ mutually exclusive sets which we can index with $v=1,...,V$. For each sample split $v = 1,...,V$, this $v^{\text{th}}$ data set represents the validation set while the complement is its training set. Let $P_{n,v}$ be the empirical distribution of the validation sample $v$, and $P_{n,-v}$ be the empirical distribution of the complementary training set. 

\subsubsection{Data and Models - Application to ``Interventions" Study}

The ``Interventions" Study is a randomized controlled trial (RCT) consisting of 441 i.i.d. observations of the following data generated by a process described by the causal model described above: covariates $W$, which includes intervention site, sex, ethnicity, age, Colorado Symptom Index (CSI) score (a measure of psychiatric symptoms), level of substance use, Level of Service Inventory (LSI) score (a measure of risk for future re-offending), number of prior adult convictions, most serious offense, Treatment Motivation Questionnaire (TMQ) score (a measure of internal motivation for undergoing treatment), and substance use level; the randomized treatment $A$, which is either a manualized Cognitive Behavioral Intervention for people criminal justice system (abbreviated CBT; $A=1$) or treatment as usual (TAU), which is mostly psychiatric or correctional services ($A=0$); and a binary outcome $Y$ of recidivism, an indicator that the person was not re-arrested over a minimum period of one year. Table \ref{table_descriptives} shows the distribution of the data.

\subsection{Causal Estimands}

In this point treatment setting, a dynamic treatment rule in the set of rules $\mathcal{D}$ is a function $d$ that takes as input some function $V$ of the measured baseline covariates $W$ and outputs a treatment decision: $V \rightarrow d(V) \in \{0,1\}$. It could be the case that $V=W$, in other words, dynamic treatment rules that potentially respond to all measured baseline covariates; in the remainder of this paper we focus on this case.

A counterfactual outcome under an arbitrary treatment rule $d$ -- an individual's outcome if, possibly contrary to fact, the individual received the treatment that would have been assigned by the treatment rule $d$ (denoted $Y_d$) -- is derived under an intervention on the above SCM. Specifically, we consider counterfactual outcomes generated by setting $A$ equal to the following treatment rules: 1) the true ODTR; and, 2) an estimate of the ODTR, either: a) the sample-specific estimate of the ODTR; or b) the training sample-specific estimate of the ODTR. 

The expectation of each of these counterfactual outcomes under the distribution $P_{U,X}$ are the causal parameters of interest in this paper. Each causal estimand is a mapping $\mathcal{M}^F \rightarrow \mathbb{R}$.

The target causal parameter corresponding to the value of a given treatment rule $d$ (from the set of rules $\mathcal{D}$) is:

\[\Psi^F_d(P_{U,X}) \equiv \mathbb{E}_{P_{U,X}} [Y_{d}].\]

\noindent The true ODTR $d_0^*$ is defined as the rule that maximizes the expected counterfactual outcome:

\[ d_0^* \in \argmax_{d \in \mathcal{D}} \Psi^F_d(P_{U,X}).\]

\noindent Here, the target causal parameter of interest is the expected outcome under the true ODTR $d_0^*$:

\[ \Psi^F_{d_0^*}(P_{U,X}) \equiv \mathbb{E}_{P_{U,X}} [Y_{d_0^*}].\]

Let $d^*_n : \mathcal{M}\rightarrow \mathcal{D}$ be an ODTR estimated on the entire sample, and $d^*_{n,v} = d^*(P_{n,-v}): \mathcal{M} \rightarrow \mathcal{D}$ be an ODTR estimated on the  $v^{th}$ training set. The data-adaptive causal parameters are: a) the expected outcome under a sample-specific estimate of the ODTR:

\[ \Psi^F_{d^*_n}(P_{U,X}) \equiv \mathbb{E}_{P_{U,X}} [Y_{d^*_n}],\]

\noindent noting that the expectation here is not over $d_n^*$, i.e., this is $\mathbb{E}_{P_{U,X}} [Y_{d}]$, evaluated at $d=d_n^*$, and b) the average of the expected validation set outcomes under training-set specific estimates of the ODTR:

\[\Psi^F_{d^*_{n,v}}(P_{U,X}) \equiv \frac{1}{V}\sum_{v=1}^V \mathbb{E}_{P_{U,X}} [Y_{d^*_{n,v}}],\]
\noindent where, the true value of this target paramter further depends on the random sample split.

One might also be interested in comparing the above causal quantities to, for example, the expected outcome had everyone been assigned the treatment $\mathbb{E}_{P_{U,X}} [Y_1 ]$ or had no one been assigned the treatment $\mathbb{E}_{P_{U,X}} [Y_0]$.

\subsubsection{Causal Estimands - Application to ``Interventions" Study}

Analagous to the above causal questions, for the ``Interventions" Study, we are interested in asking: what would have been the probability of no re-arrest had everyone been given CBT according to: 1) some pre-specified rule $d$ (for example, the simple dynamic treatment rule that gives CBT to those with a high a baseline risk score of re-offending and TAU to those with a low baseline risk score of re-offending), where the causal parameter is $\Psi_d^F(P_{U,X})$; 2) the true ODTR $d^*_0$ (the unknown dynamic treatment rule for assigning CBT that yields the highest probability of no re-arrest), where the causal parameter is $\Psi_{d^*_0}^F(P_{U,X})$; and 3) an estimate of the ODTR specific to the 441 participants in the trial, which could either be a) a sample-specific estimate $d^*_n$ (e.g., the ODTR estimated in \cite{montoyaplaceholder}) or b) a sample-split-specific estimate of the ODTR $d^*_{n,v}$? The causal parameters for a) and b) are $\Psi_{d^*_n}^F(P_{U,X})$ and $\Psi^F_{d^*_{n,v}}(P_{U,X})$, respectively.

\subsection{Identification}

Two assumptions are necessary for identification; that is, for determining that the causal estimands (a function of our counterfactual distribution) coincide with the statistical estimands (a function of our observed data distribution): the 1) randomization assumption, $ Y_a \perp A |W \text{ } a \in \{0,1\};$
and 2) positivity assumption: $ P( \min_{a \in \{0,1\}} g_0(A=a|W) > 0) = 1.$
Both hold if, for example, data are generated from an experiment in which treatment is randomized (as in the ``Interventions" trial); for data generated in an observational setting, the randomization assumption requires measurement of all unmeasured confounders, and the positivity assumption should be examined \citep{petersen2012diagnosing}.

\subsection{Statistical Estimands}

We describe statistical estimands corresponding to each of the causal parameters outlined above -- each is identified via the G-computation formula.

The statistical estimand of the mean outcome under any rule $d \in \mathcal{D}$ is

\[\psi_{0,d} \equiv \Psi_{d}(P_0) =  \mathbb{E}_0[Q_{0}(d(W),W)],\] 

\noindent where the function $Q(A,W) = \mathbb{E}[Y|A,W]$ is the outcome regression. 

\noindent The true optimal rule, as a function of the observed data distribution, is then

\[ d^*_0 \in \argmax_{d \in \mathcal{D}} \Psi_{d}(P_0).\]

\noindent Note that the RHS of this equation is a set because there may be more than one optimal rule for a certain kind of individual (e.g., if certain kinds of individuals neither benefit from nor are harmed by a treatment) \citep{luedtke2016statistical}. Here, we will assume that when there is no treatment effect, assigning treatment 0 is better than no treatment. Then, the optimal rule can  be written as a function of the so-called ``blip function", where the true blip function under $P_0$ is defined as $B_0(W) = Q_0(1,W) - Q_0(0,W)$:

\[ d^*_0(W) = \mathbb{I}[B_0(W) > 0].\]

\noindent The true mean outcome under the true optimal rule $d_0^*$ is then identified by

\[\psi_{0,d^*_0} \equiv \Psi_{d^*_0}(P_0) = \mathbb{E}_0[Q_{0}(d^*_0(W),W)].\]

\noindent The first data-adaptive parameter we consider, as a function of the observed data, is the true expected outcome under an ODTR estimate based on the entire sample $d^*_n$:

\[\psi_{0,d^*_n} \equiv \Psi_{d^*_n}(P_0) = \mathbb{E}_0[Q_{0}(d^*_n(W),W)].\]

\noindent The second data-adaptive parameter is the average of the validation-set true mean outcomes under the training-set estimated ODTRs $d^*_{n,v}$:

\[\psi_{0,d^*_{n,v}} \equiv \Psi_{d^*_{n,v}}(P_0) =  \frac{1}{V}\sum_{v=1}^V\mathbb{E}_0[Q_{0}(d^*_{n,v}(W),W)].\]

\section{Estimation}

We describe estimators for each of the statistical parameters above: a simple substitution estimator based on the G-computation formula, an IPTW estimator, a double-robust IPTW estimator (IPTW-DR), a TMLE, and a CV-TMLE. Each of these estimators can be used for estimating $\psi_{0,d}$ and $\psi_{0,d_0^*}$. We use the non-cross-validated estimators (G-computation, IPTW, IPTW-DR, and TMLE) to estimate $\psi_{0,d_n^*}$; we estimate $\psi_{0,d^*_{n,v}}$ with CV-TMLE. 

%When employing these estimators to approximate $\psi_{d_0^*}$, we examine the cases in which the true ODTR $d_0^*$ is known (in order to illustrate performance of these estimators when evaluating any true, known rule) and unknown (in order to illustrate the realistic scenario in which the ODTR needs to be both estimated and evaluated).

Let $Q_n$ be an estimator of the outcome regression, which could be estimated with, for example, SuperLearner \citep{van2007super}. In a randomized experiment, the treatment mechanism $g_0$ is known; thus, one could use this known $g_0$, or $g_n$ could be a maximum likelihood estimator (MLE) based on a correctly specified model. 

\subsection{Non-cross-validated estimators for estimating $\psi_{0,d}$, $\psi_{0,d_0^*}$, and $\psi_{0,d_n^*}$}

We first illustrate each of the non-cross-validated estimators suited for estimating a treatment-specific mean at an arbitrary $d \in \mathcal{D}$, which, for example, could be an \textit{a priori} known rule or an optimal rule estimated on the entire sample (see \cite{luedtkeSLODTR} and \cite{montoyaplaceholder} for a description on how to estimate the optimal rule using, for example, the ODTR SuperLearner). Here, $\hat{\Psi}_d(P_n) \equiv \hat{\psi}_d$ is an estimate of the true parameter value $\psi_{0,d}$, based on applying the estimator $\hat{\Psi}_d$ to an empirical distribution based on sampling from $P_0$. We further subscript by each estimator name.

One can use a(n):
\begin{itemize}
    \item Simple substitution estimator based on the above G-computation formula,
\[\hat{\psi}_{gcomp,d}=\frac{1}{n} \sum_{i=1}^n Q_n(d(W_i),W_i);\]
\item IPTW estimator,
\[\hat{\psi}_{IPTW,d}=\frac{1}{n} \sum_{i=1}^n \frac{\mathbb{I}[A_i=d(W_i)]}{g_n(A_i|W_i)}Y_i;\]
\item Double-robust IPTW estimator,
\[\hat{\psi}_{IPTW-DR,d}=\frac{1}{n} \sum_{i=1}^n \left[\frac{\mathbb{I}[A_i=d(W_i)]}{g_n(A_i|W_i)}(Y_i-Q_n(A_i,W_i)) + Q_n(d(W_i),W_i)\right];\]
\item or TMLE. We briefly describe one possible TMLE procedure. First, estimate the $i$-specific so-called clever covariate:

\[H_{n,i} = \frac{\mathbb{I}[A_i=d(W_i)]}{g_n(A_i|W_i)}.\]

\noindent Then, update the initial fit of $Q_n$ by running a logistic regression of $Y$ (which should be transformed between $0$ and $1$ if the outcome is continuous \citep{gruber2010targeted}) on offset $Q_n(d(W),W)$ with weights $H_{n}$, with maximum likelihood estimation used to estimate the intercept. Denote the predictions from this logistic regression as $Q_n^* (d(W),W)$, from the updated fit. Then, the TMLE estimator is:

\[\hat{\psi}_{TMLE,d}=\frac{1}{n} \sum_{i=1}^n  Q_n^*(d(W_i),W_i).\]
\end{itemize}

\subsection{CV-TMLE for estimating $\psi_{0,d}$, $\psi_{0,d_0^*}$, and $\psi_{0,d^*_{n,v}}$}

As previously mentioned, the CV-TMLE can estimate $\psi_{0,d}$, $\psi_{0,d_0^*}$, and $\psi_{0,d^*_{n,v}}$. Instead of illustrating the CV-TMLE at $d$ as in the above estimators, we illustrate one type of CV-TMLE procedure for evaluating the mean outcome under sample-split-specific estimates of the ODTR $d^*_{n,v}$ to show on which parts of the data one needs to estimate or predict the ODTR, if estimating $\psi_{0,d_0^*}$ or $\psi_{0,d^*_{n,v}}$. The same procedure holds for a $d$ that is known, except that the rule need not be estimated on each of the training samples and is simply applied to the validation sets:

\begin{enumerate}
    \item Split the data into $V$ folds. Let each fold be the validation set and the complement data be the training set, providing us with $v$-specific sample splits in the validation and training set.
    \item For $v=1,...,V$, carry out the following steps:
    \begin{enumerate}
    \item Estimate the treatment mechanism, ODTR, and outcome regression on the training set. 
    \item Using the fits from the previous step, generate predictions of the observed treatment assignment, optimal treatment assignment, and outcome under the optimal treatment assignment for observations in the validation set. Denote the corresponding estimates as $g_{n,v}(A|W)$, $d_{n,v}^*(W)$, and $Q_{n,v}(A=d_{n,v}^*(W),W)$.
    \item Update $Q_{n,v}$ generated in the previous step by fitting an intercept model (as described in the TMLE updating procedure in the previous subsection) on persons in the validation set. Call the updated fit $Q^*_{n,v}$.
    \item Generate validation set-specific targeted estimates of the mean outcome under the sample-split-specific estimated rule $d^*_{n,v}$ by evaluating $Q^*_{n,v}$ on data in the validation set. Call the updated estimates $Q^*_{n,v}(A = d^*_{n,v},W)$.
    \item Define the $v^{\text{th}}$ validation set-specific estimate of the mean outcome under the estimate rule as: 
    \[\hat{\psi}_{d^*_{n,v}} = \frac{1}{n_v}\sum_{i\in Val(v)}Q^*_{n,v}(d^*_{n,v}(W_i),W_i)),\]
    where $n_v$ denotes the number of individuals in the validation set $v$ and $Val(v)$ is the indices $i$ for which $O_i$ is in the validation set.
    %Using the predictions from the previous step, in each corresponding validation set, update the initial estimator $\hat{\Psi}_{d^*_{n,v}}(P_{n,-v})$ that uses $Q_{n,v}$ via the TMLE procedure described above to generate $\hat{\Psi}_{d^*_{n,v}}(P^*_{n,-v})$ that uses $Q^*_{n,v}$ (the update of $Q_{n,v}$)}, a TMLE of $\mathbb{E}_0[Q_0(d^*_{n,v}(W),W)]$.
    \end{enumerate}
    \item Average over all validation folds to obtain the CV-TMLE, i.e., the estimated mean outcome under the sample-split-specific estimates of ODTR:
    
    \[\hat{\psi}_{CV-TMLE,d^*_{n,v}} =  \frac{1}{V}\sum_{v=1}^{V}\hat{\psi}_{d^*_{n,v}}.\]
\end{enumerate}

\section{Inference}

We first discuss the conditions necessary for each the above estimators to be asymptotically linear for $\psi_{0,d}$, $\psi_{0,d_n^*}$, and $\psi_{0,d^*_{n,v}}$ in a randomized experiment.  Under these conditions, using influence-curve based inference, we describe how to construct 95\% confidence intervals with nominal to conservative coverage for the aforementioned statistical estimands of interest.

We do not discuss inference on the G-computation estimator, because in order for it to be asymptotically linear, $Q_n$ must either be equal to $Q_0$ or be an estimator that converges fast enough to $Q_0$, neither of which we assume here. 

For more details and proofs, we refer the reader to \cite{TLBBD}, \cite{van2015targeted}, and \cite{van2003unified}.

\subsection{Asymptotic Linearity Conditions for Estimators}

We give a brief overview of the conditions needed for asymptotic linearity for each of the estimators with respect to each statistical estimand in the randomized trial setting, and provide an informal summary of these conditions in Table \ref{table0}. 

An estimator $\hat{\Psi}$ is asymptotically linear for its true value $\psi_0$ if it can be written in the following form:

\[\hat{\psi} - \psi_0 = \frac{1}{n}\sum_{i=1}^n IC(O_i) + R_n,\]

\noindent where $IC$ is the estimator's influence curve (that is centered to have mean $0$, by definition) and $R_n$ is a remainder term that is $o_P(1/\sqrt{n})$. An asymptotically linear estimator $\hat{\Psi}$ thus generally has the following properties: 1) its bias converges to 0 in sample size at a rate faster than $\frac{1}{\sqrt{n}}$; 2) for large $n$, its distribution is approximately normal, $n^{1/2}(\hat{\psi} - \psi_0) \overset{d}{\to} N(0, \sigma^2_0)$, allowing an estimate of $\sigma^2_0$ to be used to construct a Wald-type confidence intervals; and, 3) the asymptotic variance of $n^{1/2}(\hat{\psi} - \psi_0)$ (i.e., $\sigma^2_0$) can be well-approximated by the sample variance of its estimated influence curve $IC_n$ (or equivalently, $\sigma^2_n = \frac{1}{n}\sum_{i=1}^n IC^2_n(O_i)$, since the mean of an influence curve is 0).

\subsubsection{Conditions for value of a known rule}

Our randomized experiment scenario guarantees that $g_0$ is known, and thus $g_n$ can be a maximum likelihood estimate (MLE) of $g_0$  based on a correctly specified parametric model. As a result, for an estimand defined as the value of an \textit{a priori} specified rule $d$, the IPTW estimator is guaranteed to be asymptotically linear for $\psi_{0,d}$; however, this estimator will not be asymptotically efficient. 

Let $Pt = \mathbb{E}_P[t(O)]$ for a distribution $P$ and function $t$, and let $IC^*$ be the efficient influence curve. If $P_0\{IC^*(Q_n,g_n)-IC^*(Q,g_0)\}^2$ converges to zero in probability for a limit $Q$, possibly misspecified, and
$\{IC^*(Q,g_0):Q\}$ is a $P_0$ Donsker class, such as the class of $d$ variate cadlag functions with a universal bound on the sectional variation norm, then the TMLE and IPTW-DR are asymptotically linear with an influence curve equal to $IC^*(Q,g_0)$ minus its projection onto the tangent space of the parametric model used for the MLE $g_n$. As a consequence, its asymptotic variance is smaller than or equal to the variance of $IC^*(Q,g_0)$, and, in particular, if $g_n$ is replaced by $g_0$, then $IC^*(Q,g_0)$ is the actual influence curve. Further, if $Q_n$ is consistent for $Q_0$ in the sense that $P_0\{IC(Q_n,g_0)-IC(Q_0,g_0)\}^2$ converges to zero in probability, then the TMLE and IPTW-DR estimators are also asymptotically efficient. 

The above is also true for CV-TMLE, except Donsker class conditions can now be removed (in effect allowing for an overfit in the initial estimate of $Q_0$).

\subsubsection{Conditions for value of true ODTR}

Construction of nominal to conservative confidence intervals around each of the non-cross-validated estimators with respect to the true expected outcome under the true, unknown $d^*_0$ requires additional assumptions. For these estimators, statistical inference for $\psi_{0,d^*_0}$ relies on a second-order difference in $R_n$ between $\psi_{0,d^*_n}$ and $\psi_{0,d^*_0}$ going to 0 at a rate faster $1/\sqrt{n}$. In practice, how hard it is to make this condition hold depends on the extent to which the blip function as a random variable (function of $W$) has density at zero. If the value of the blip is always larger than $\delta > 0$ for some $\delta > 0$, then consistency of $Q_n$ is sufficient; however, if the treatment effect is zero for some covariate values that have positive probability of occurring, then stronger assumptions are required \citep{luedtkeSLODTR, van2015targeted}. The non-cross-validated estimators additionally require Donsker class conditions on $IC_{d^*_n}(Q_n,g_n)$, thereby also restricting the adaptivity of $d_n^*$ (informally, that $d^*_n$ not be an overfit of $d_0^*$). In practice, these conditions on the data-adaptivity of $d^*_n$ hold if, for example, the optimal rule is a function of one covariate, or, if a higher-dimensional covariate set is used, one is willing to make strong smoothness assumptions, for example, on the blip function. 

CV-TMLE also relaxes these additional Donsker conditions on $d^*_n$. Thus, in a randomized trial, if employing CV-TMLE for $\psi_{0,d^*_0}$, the only condition needed is that $\psi_{0,d_{n,v}^*} - \psi_{0,d^*_0} = o_P(n^{-1/2})$. This condition is carefully addressed in \cite{luedtkeSLODTR} and can be expressed in terms of a condition on $P(B_0<x)$ to converge to zero at a fast enough rate in $x$ as $x$ approaches $0$.

\subsubsection{Conditions for value of sample-(split)-specific ODTR estimate}
For the data-adaptive parameters, the asymptotic study of the non-cross-validated estimators no longer requires the strong assumption that $d^*_n$ converges to $d^*_0$ at a fast enough rate; rather, they only require that $d^*_n$ converges to some fixed rule $d \in \mathcal{D}$ at any rate. 

Similarly, CV-TMLE only requires the weak consistency condition that $d^*_{n,v}$ converges to some fixed rule $d \in \mathcal{D}$ at any rate. This means that, for randomized trial data, and under the above analogue $L^2(P_0)$ consistency conditions, the CV-TMLE estimator for $\psi_{0,d^*_{n,v}}$ is asymptotically linear under essentially no conditions for the data adaptive parameter $\psi_{0,d^*_{n,v}}$.

%Specifically, for the CV-TMLE of the data dependent target parameter we have that, under the above analogue $L^2(P_0)$ consistency conditions, $\psi_{CV-TMLE,d^*_{n,v}} - \psi_{0,d^*_{n,v}}$ behaves as $\frac{1}{V} \sum_v P_{n,v} IC^*_{d_{n,v}}(Q,g_0)$ and is thereby asymptotically normal with variance
%$\frac{1}{V}\sum_v \sigma^2_v$, where $\sigma^2_v$ is the limit of the variance of $IC^*_{d_{n,v}}(Q,g_0)$ (treating the training sample as given). 

%Under a weak consistency condition on $d_{n,v}$, the CV-TMLE is asymptotically linear with influence curve $IC^*_{d}(Q,g_0)$ with $d$ being a limit of $d_{n,v}$, such as possibly the optimal rule, in which case it is asymptotically linear with influence curve $IC^*_{d_0}(Q,g_0)$. 

\begin{center}
\begin{table}[htb]
\scalebox{.75}{
\begin{tabular}{|l|l|c|l|l|l|}
\hline
\multicolumn{2}{|l|}{} & \multicolumn{4}{c|}{\textbf{Conditions for Asymptotic Linearity:}} \\ \hline
\textbf{Estimands} & \textbf{Estimators} & \multicolumn{1}{l|}{\textbf{\begin{tabular}[c]{@{}l@{}}$g_n = g_0$ or $g_n$ is MLE \\ of $g_0$ according to \\ correctly specified\\ parametric model\end{tabular}}} & \textbf{\begin{tabular}[c]{@{}l@{}}Fast enough \\ convergence of \\ $d_n^*$ to $d^*_0$\end{tabular}} & \textbf{\begin{tabular}[c]{@{}l@{}}$Q_n$ not \\ overfit\end{tabular}} & \textbf{\begin{tabular}[c]{@{}l@{}}$d_n$ not \\ overfit\end{tabular}} \\ \hline
\multirow{4}{*}{\textbf{\begin{tabular}[c]{@{}l@{}}Value of \\ known rule \\ $\psi_{0,d}$\end{tabular}}} & $\hat{\Psi}_{IPTW,d}$ & \multirow{12}{*}{\begin{tabular}[c]{@{}c@{}}Satisfied by \\ randomized \\ experiment\end{tabular}} & \multirow{4}{*}{\begin{tabular}[c]{@{}l@{}}Not required, \\ $d$ known\end{tabular}} & Not required & \multirow{4}{*}{\begin{tabular}[c]{@{}l@{}}Not required, \\ $d$ known\end{tabular}} \\ \cline{2-2} \cline{5-5}
 & $\hat{\Psi}_{IPTW-DR,d}$ &  &  & Required &  \\ \cline{2-2} \cline{5-5}
 & $\hat{\Psi}_{TMLE,d}$ &  &  & Required &  \\ \cline{2-2} \cline{5-5}
 & $\hat{\Psi}_{CV-TMLE,d}$ &  &  & Not required &  \\ \cline{1-2} \cline{4-6} 
\multirow{4}{*}{\textbf{\begin{tabular}[c]{@{}l@{}}Value of \\ true ODTR \\ $\psi_{0,d_0^*}$\end{tabular}}} & $\hat{\Psi}_{IPTW,d^*_n}$ &  & \multirow{3}{*}{Required} & Not required & \multirow{3}{*}{Required} \\ \cline{2-2} \cline{5-5}
 & $\hat{\Psi}_{IPTW-DR,d^*_n}$ &  &  & Required &  \\ \cline{2-2} \cline{5-5}
 & $\hat{\Psi}_{TMLE,d^*_n}$ &  &  & Required &  \\ \cline{2-2} \cline{4-6} 
 & $\hat{\Psi}_{CV-TMLE,d^*_{n,v}}$ &  & Required & Not required & Not required \\ \cline{1-2} \cline{4-6} 
\multirow{3}{*}{\textbf{\begin{tabular}[c]{@{}l@{}}Value of sample-\\ specific ODTR \\ estimate $\psi_{0,d_n^*}$\end{tabular}}} & $\hat{\Psi}_{IPTW,d^*_n}$ &  & \multirow{3}{*}{\begin{tabular}[c]{@{}l@{}}Not required; \\ require $d^*_n \overset{p}{\to} d \in \mathcal{D}$\end{tabular}} & Not required & \multirow{3}{*}{Required} \\ \cline{2-2} \cline{5-5}
 & $\hat{\Psi}_{IPTW-DR,d^*_n}$ &  &  & Required &  \\ \cline{2-2} \cline{5-5}
 & $\hat{\Psi}_{TMLE,d^*_n}$ &  &  & Required &  \\ \cline{1-2} \cline{4-6} 
\textbf{\begin{tabular}[c]{@{}l@{}}Value of \\ sample-split-specific \\ ODTR estimate \\ $\psi_{0,d_{n,v}^*}$\end{tabular}} & $\hat{\Psi}_{CV-TMLE,d^*_{n,v}}$ &  & \begin{tabular}[c]{@{}l@{}}Not required; \\ require $d^*_{n,v} \overset{p}{\to} d \in \mathcal{D}$\end{tabular} & Not required & Not required \\ \hline
\end{tabular}
}
%\end{adjustbox}
\caption{Informal summary of the conditions needed for asymptotic linearity in the randomized treatment setting for each of the estimators corresponding to each of the estimands. }
\label{table0}
\end{table}
\end{center}

\subsection{Construction of Confidence Intervals}

Below, we list conservative working influence curves for each estimator at $P_n$ and $d \in \mathcal{D}$. The actual estimators' influence curves when an MLE of $g_n$ based on a correctly specified parametric model is used (as can be guaranteed when treatment is randomized) are the working influence curves presented below minus a tangent space projection term. Thus, under the conditions stated above, the sample variance of the following working influence curves at a correctly specified $g_n$ yield conservative estimates of the asymptotic variance of the estimators, which yields conservative confidence interval coverage.

%.  This means that the confidence intervals generated with the following influence curves evaluated at $g_0$ will yield nominal coverage; the influence curves at a correctly specified parametric model for $g_n$ (as presented below) will yield 

%The influence curves of each estimator, if instead evaluated at the true $g_0$, equal the influence curves listed below.

The IPTW estimator's working influence curve estimate is:

\[ \widehat{IC}_{IPTW,d}(O) = \frac{\mathbb{I}[A=d]}{g_n(A|W)}Y  - \hat{\psi}_{IPTW,d}.\]

\noindent The influence curve of the TMLE and double-robust IPTW estimator is the \emph{efficient} influence curve for the treatment-specific mean \citep{van2000asymptotic, bickel1993efficient}; the corresponding working influence curve estimates are:

\[ \widehat{IC}_{IPTW-DR,d}(O) = \frac{\mathbb{I}[A=d]}{g_n(A|W)} (Y - Q_n(A,W)) + Q_n(d(W),W)  - \hat{\psi}_{IPTW-DR,d},\]

\[ \widehat{IC}_{TMLE,d}(O) = \frac{\mathbb{I}[A=d]}{g_n(A|W)} (Y - Q^*_n(A,W)) + Q^*_n(d(W),W)  - \hat{\psi}_{TMLE,d}.\]

As stated above, for these non-cross-validated estimators, the asymptotic variance can be conservatively estimated with the sample variance of the estimated influence curve: $\sigma^2_n = \frac{1}{n}\sum_i^n \widehat{IC}^2(O_i)$.

For the IPTW-DR and TMLE estimators, one can underestimate the estimator's variance if $Q_0$ is estimated data-adaptively on the same data on which the sample variance of the estimated influence curve is evaluated. Through sample splitting, CV-TMLE confidence intervals protect against overfitting incurred by using the data twice -- for both estimation and evaluation. Then the fold-specific estimate of the working influence curve for CV-TMLE is based on estimating $d^*_{0}$, $Q_{0}$, and $g_{0}$ on the $v^{\text{th}}$ training sample, evaluated on the complementary validation sample:

\[\widehat{IC}_{v,d^*_{n,v}}(O) = \frac{\mathbb{I}[A=d^*_{n,v}(W)]}{g_{n,v}(A|W)} (Y- Q^*_{n,v}(A,W)) + Q^*_{n,v}(d^*_{n,v}(W),W)  - \hat{\psi}_{d^*_{n,v}},\]

\noindent and the fold-specific estimate of the variance of the fold-specific estimator is:

\[\sigma^2_{n,v} = \frac{1}{n_v}\sum_{i\in v}\widehat{IC}_{v,d^*_{n,v}}^2(O_i),\]
where, as before, $n_v$ denotes the number of individuals in validation set $v$; thus, the asymptotic variance of the CV-TMLE $\hat{\psi}_{CV-TMLE,d^*_{n,v}}$ can be conservatively estimated with:

\[\sigma^2_{n,CV-TMLE} = \frac{1}{V}\sum_{v=1}^{V}\sigma^2_{n,v}.\]

In sum, for each estimator $\hat{\Psi}$ and its corresponding working influence curve estimate $IC_n$, we obtain conservative inference on the value of the rule by constructing confidence intervals in the following way:

\[\hat{\psi} \pm \Phi^{-1}(0.975)\frac{\sigma_n}{\sqrt{n}}.\]

\section{Simulation Study}

Using simulations, we evaluate the performance of various estimators of the value of the rule in finite samples. In particular, we investigate: 1) the impact of increasingly data-adaptive estimation of nuisance parameters and (where applicable) the ODTR; 2) the potential for efficiency and bias improvement through the use of semiparametric efficient estimators; and, 3) the importance of sample splitting, in particular via a cross-validated-targeted maximum likelihood estimator (CV-TMLE). 

\subsection{Data Generating Process}

All simulations were implemented in R \citep{R}, and the code, simulated data, and results can be found at https://github.com/lmmontoya/SL.ODTR. In the future, we plan to integrate the SL.ODTR software to the Targeted Learning software ecosystem, tlverse \citep{tlverse}. We examine these comparisons using the following data generating process (DGPs) (also used in \cite{montoyaplaceholder, van2015targeted, luedtkeSLODTR}). Each simulation consists of 1,000 iterations of $n$=1,000 observations. Mimicking a randomized experiment, the (independent) covariates, treatment and outcome are generated as follows:
\begin{align*}
W_1,W_2,W_3,W_4 \sim & Normal(\mu=0,\sigma^2=1), \\
A \sim & Bernoulli(p=0.5), \\
Y \sim & Bernoulli(p=Q_0(A,W)),\\
\text{where } Q_0(A,W) =& 0.5\textrm{expit} (1-W_1^2  + 3W_2  + 5W_3^2 A - 4.45A)+ \\
& 0.5\textrm{expit} (-0.5- W_3  + 2W_1 W_2  + 3|W_2|A - 1.5A),
\end{align*}
then the true blip function is:
\begin{align*}
    B_0 (W)= & 0.5[\textrm{expit} (1-W_1^2  + 3W_2  + 5W_3^2  - 4.45)+\textrm{expit} (-0.5- W_3  + 2W_1 W_2  + 3|W_2|  - 1.5)\\
& - \textrm{expit} (1-W_1^2  + 3W_2 )-\textrm{expit} (-0.5- W_3  + 2W_1 W_2 )].
\end{align*}
Here, the true expected outcome under the true ODTR $\Psi^F_{d_0^*}(P_{U,X}) \approx 0.5626$ and the true optimal proportion treated $\mathbb{E}_{P_{U,X} } [d_0^* ] \approx 55.0\%$. The mean outcome had everyone and no one been treated are, respectively, $\mathbb{E}_{P_{U,X}} [Y_1 ] \approx 0.4638$ and $\mathbb{E}_{P_{U,X}} [Y_0 ] \approx 0.4643$.

In the Appendix, we illustrate results on the same simulation procedure, but with dependent covariates.

\subsection{Estimator Configurations}

We estimate each of the statistical estimands using the IPTW, IPTW-DR, TMLE, and CV-TMLE estimators, with inference based on the conservative working influence curves describe above. The G-computation estimator is also employed, but confidence intervals are not generated.

A correctly specified logistic regression is used to estimate the nuisance parameter $g_0$, reflecting the RCT setting. SuperLearner is used to estimate $Q_0$ and $d^*_0$. The ODTR is estimated using a ``blip-only" library, using a blip-based metalearner (i.e., an approach to creating an ensemble of candidate ODTR algorithms), and using the value of the candidate rule as the risk function \citep{montoyaplaceholder}. Three libraries are considered that correspond to varying levels of data-adaptiveness, or potential for overfitting.

\begin{enumerate}
    \item ``GLMs - least data adaptive"
    \begin{itemize}
        \item $Q$ library: four logistic regressions, each with a main terms $W_j$ and $A$, and with an interaction $W_j$ times $A$, for $j \in \{1,..,4\}$
        \item $d^*$ library: univariate linear regressions with each covariate
    \end{itemize}
    \item ``ML + GLMs - moderately data adaptive"
    \begin{itemize}
        \item $Q$ and $d^*$ library: all algorithms in the ``GLMs - least data adaptive" $Q$ and $d^*$ libraries, respectively, in addition to the algorithms \texttt{SL.glm} (generalized linear models), \texttt{SL.mean} (the average), \texttt{SL.glm.interaction} (generalized linear models with interactions between all pairs of variables), \texttt{SL.earth} (multivariate adaptive regression splines \citep{friedman1991multivariate}), \texttt{SL.nnet} (neural networks \citep{ripley1996pattern}), \texttt{SL.svm} (support vector machines \citep{chang2011libsvm}), and \texttt{SL.rpart} (recursive partitioning and regression trees \citep{breiman2017classification}) from the SuperLearner package \citep{SLpackage}
    \end{itemize}
    \item ``ML + GLMs - most data adaptive"
    \begin{itemize}
        \item $Q$ and $d^*$ library: all algorithms in the ``ML + GLMs - moderately data adaptive" $Q$ and $d^*$ libraries, respectively, in addition to \texttt{SL.randomForest} \citep{breiman2001random}
    \end{itemize}
\end{enumerate}

\subsection{Performance Metrics}

Using measures of bias, variance, mean squared error (MSE) and 95\% confidence interval coverage, we evaluate the ability of each of the estimators to approximate: 1) the true expected outcome under an \textit{a priori} known rule $d$, i.e., $\psi_{0,d}$; 2) the true expected outcome under the true, unknown ODTR $\psi_{0,d_0^*}$; 3) the true expected outcome under an ODTR estimated on: a) the entire sample and evaluated on the entire sample $\psi_{0,d^*_n}$; or b) estimated on each of the training sets, evaluated and averaged over each of the validation sets $\psi_{0,d^*_{n,v}}$. 

First, we estimate the target parameter $\psi_{0,d}$. This illustrates the performance of these estimators of the value of a rule when the rule is known \textit{a priori}, either because the rule is known to be of interest or it was estimated on other data not included in the current sample. In this case, we choose $d$ to be the true ODTR, that is, $d = d^*_0$. We note that it is highly unlikely that in practice $d^*_0$ is known \textit{a priori}, and stress that the only reason we examine the performance of estimators $\hat{\psi}_{d=d^*_0}$ with respect to $\psi_{0,d^*_0}$ is to illustrate how well these estimators evaluate a given pre-specified rule. However, illustrating this using the true rule $d^*_0$ in a simulation facilitates comparison of estimator performance across estimands, showing, for example, the price in performance one pays for targeting the more ambitious parameter that seeks to estimate both the rule itself and its true value. Said another way, if we see that estimator performance for $\hat{\psi}_{d=d^*_0}$ with respect to $\psi_{0,d^*_0}$ is good, then the only issue left with estimating $\psi_{0,d^*_0}$ is estimating $d^*_0$ well.

Next, we estimate the same target parameter $\psi_{0,d^*_0}$ in the more realistic scenario where the true ODTR $d^*_0$ is unknown. We therefore first estimate the ODTR and then apply each of the estimators of the value of the rule under the estimated ODTR (where the rule is either estimated on the entire sample $\hat{\psi}_{d^*_n}$ or, for CV-TMLE, estimated on each sample split $\hat{\psi}_{d^*_{n,v}}$). Performance of the estimators with respect to $\psi_{0,d^*_0}$ reflects how well both the rule and its value are estimated.

Finally, we treat as target parameter the true expected outcome under the estimated optimal rule, i.e., the data-adaptive parameters $\psi_{0,d^*_n}$ or, for CV-TMLE, $\psi_{0,d^*_{n,v}}$. This illustrates estimator performance for data-adaptive parameters whose true values depend on the sample, and for which it is of interest to estimate their value using the same sample on which the rule was learned. Note that the target parameter value in this case is specific to the sample at hand (the ``truth" will vary from sample to sample); thus, performance calculations are calculated with respect to the true sample-specific or sample-split-specific mean outcome. For example, for confidence interval coverage, across the 1,000 simulations, we calculated the proportion of times the confidence interval around the estimated value of the estimated rule covered the true value of the estimated rule -- where both the confidence interval around the estimate and the true value of the estimated rule are \emph{specific to each sample}. Furthermore, the data-adaptive parameter will vary between the non-cross-validated estimators (whose data-adaptive parameter is the sample-specific parameter $\psi_{0,d^*_n}$) and CV-TMLE (whose data-adaptive parameter is the sample-split-specific parameter $\psi_{0,d^*_{n,v}}$), and as such, is not only a function of the sample, but also of the split.

\subsection{Simulation Results}

%Figure \ref{plotfig} and Tables \ref{table1}, \ref{table2}, and \ref{table3} display simulation results. Below we discuss results specific to each metric of estimator performance, library configuration corresponding to the degree of data-adaptivity for estimating the ODTR and nuisance parameters, and estimator for the value of the rule. 

\subsubsection{Results - Value of a Known Dynamic Treatment Regime}

Bias, variance, MSE, and confidence interval coverage metrics for estimating $\psi_{0,d}$ in the scenario where $d$ is known \textit{a priori} illustrate the performance of each of the estimators for estimating the value of a given pre-specified rule. For illustration, we use the true optimal rule $d^*_0$. Thus, only estimation of nuisance parameters $g$ and/or $Q$ were needed for this parameter.

The untargeted G-computation formula exhibited considerable bias if either misspecified parametric models or a SuperLearning approach was used to estimate the outcome regression -- regardless of the degree of data-adaptiveness in estimating this nuisance parameter $Q$. For example, when the $Q_n$ library consisted of only parametric regressions, the mean difference between the G-computation estimate and the truth was $-9.35\%$ (i.e., 133.57-467.50 times that of the bias of alternative estimators). We note that this result is in contrast to that of estimating the treatment specific mean for any static regime (in which treatment assignment is not a function of covariates, for example, $\mathbb{E}_0[Q_0(A=1,W)]$) from data generated from a randomized experiment. In the latter, the G-computation estimator under certain misspecified parametric models is a TMLE, and is therefore unbiased \citep{rosenblum2009using}. 

As expected, the IPTW estimator, although unbiased, was less efficient than alternative estimators -- specifically, the IPTW estimator's variance was 1.40-2.00 times that of the variance of double-robust estimators. Additionally, the IPTW-DR and TMLE were unbiased (as expected, given the double-robustness of these estimators) if the outcome regression was estimated using either a regression based on a misspecified parametric model or a SuperLearner with a less data-adaptive library. However, both estimators were biased (i.e., $-0.74\%$ and $-0.67\%$ bias for IPTW-DR and TMLE, respectively) with less than nominal confidence interval coverage (i.e., $90.5\%$ and $91.0\%$ coverage for IPTW-DR and TMLE, respectively) when a more data-adaptive library was used to estimate the outcome regression -- a result likely due to overfitting $Q_n$. 

Sample-splitting via CV-TMLE removed the non-cross-validated estimators' bias ($-0.07\%$, or 0.006-0.100 times the bias relative to alternative double-robust estimators) and generated better confidence interval coverage ($94.7\%$) under the presence of overfitting for $Q_n$, at no cost to variance. 

\subsubsection{Results - Value of the True, Unknown ODTR}

No estimator performed well when both the ODTR itself and its value were estimated using the same sample (i.e., estimators $\hat{\psi}_{d^*_n}$ or $\hat{\psi}_{d^*_{n,v}}$ for $\psi_{0,d^*_0}$). This was evident particularly in terms of increased bias when a less data-adaptive library was used to estimate $Q_0$ and $d^*_0$, and in terms of both increased bias and variance when a more aggressive library was used to estimate $Q_0$ and $d^*_0$. Notably, however, CV-TMLE performed the best with respect to all performance metrics under the most data-adaptive approaches. A large component of the bias in this case was due to the rate of convergence from $d^*_n$ to $d^*_0$ for any SuperLearner library. As a result, confidence interval coverage of the true value under the true ODTR around any estimated value of the estimated rule did not approach 95\% (confidence interval coverage under the least, moderately, and most data adaptive libraries ranged from 14.00\%-45.70\%, 69.90\%-79.50\%, and 38.30\%-71.90\%, respectively).

Although the focus of these simulations was not optimizing estimation of the ODTR, we note that, consistent with results from \cite{montoyaplaceholder}, the least biased estimators of the true value of the true ODTR are ones that use a combination of regressions based on parametric models and machine learning algorithms in the estimation of $Q_0$ and $d_0$.% (for a thorough discussion of best practices for estimating the ODTR SuperLearner we refer the readers to Montoya, et. al. \citep{montoyaplaceholder}). 

\subsubsection{Results - Value of an Estimated ODTR}

We evaluated the performance of the non-cross-validated estimators (IPTW, IPTW-DR, and TMLE, i.e., $\hat{\psi}_{d^*_n}$) of the data-adaptive parameter (i.e., $\psi_{0,d^*_n}$) -- a parameter that depends on the optimal rule specific to the sample at hand. All non-cross-validated estimators overestimated the value of the rule (i.e., positive bias), regardless of the SuperLearner library. In addition, the bias increased as the library for estimating the ODTR became more data-adaptive. For example, for the most data-adaptive SuperLearner library configuration, TMLE exhibited a bias of 12.74\%, variance of 0.0118, MSE of 0.0280, and 22.50\% confidence interval coverage.

The CV-TMLE (i.e., $\hat{\psi}_{CV-TMLE, d^*_{n,v}}$) with respect to the data-adaptive parameter $\psi_{0,d^*_{n,v}}$ removed the bias incurred by estimating and evaluating the ODTR on the same sample, at little cost to no cost to variance. For example, for the most data-adaptive SuperLearner library configuration, CV-TMLE had a bias of 0.13\% (0.0090-0.0154 times that of alternative estimators), variance of 0.0007 (0.06-1.00 times that of alternative estimators), MSE of 0.0007 (0.02-0.09 times that of alternative estimators), and 93.9\% confidence interval coverage.

\section{Evaluating the Estimated ODTR for the ``Interventions" Study}

In our companion paper, we estimated the ODTR on the ``Interventions" data ($n=441$) using the ODTR SuperLearner. The library for $d^*_n$ consisted of a combination of algorithms based on simple parametric models and machine learning (\texttt{SL.glm}, \texttt{SL.mean}, \texttt{SL.glm.interaction}, \texttt{SL.earth}, and \texttt{SL.rpart}), and we used the same library for $Q_n$. The ODTR algorithm allocated all coefficient weight on a simple GLM with only substance use; this means that the estimated ODTR can be interpreted as: give CBT to those with low substance use scores and TAU to those with high substance use scores. 

In this paper, we \emph{evaluate} this estimated ODTR using CV-TMLE. Specifically, we aim to determine if administering CBT under this individualized rule is better than administering CBT in a non-individualized way -- i.e., simply giving all participants CBT or no participants CBT.

\begin{figure}[h]
    \centering
    \includegraphics[width=1\textwidth]{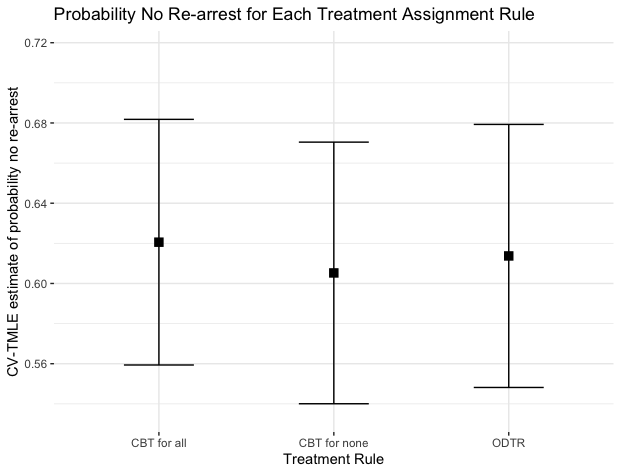}
    \caption{Analysis of the ``Interventions" Study. CV-TMLE estimates of the probability of no re-arrest under the following treatment rules: give cognitive behavioral therapy (CBT) to all, give CBT to none, give CBT according to the ODTR SuperLearner algorithm. The squares are the point estimates and the error bars are 95\% confidence intervals on these point estimates. There is no significant difference in the estimated probability of no re-arrest under a treatment regime in which all are given CBT, none are given CBT, and CBT is given using this ODTR.}
    \label{eydopt}
\end{figure}

The CV-TMLE estimate of the probability of no re-arrest under the ODTR SuperLearner is 61.37\% (CI: [54.82\%, 67.93\%]). However, this probability is not significantly different than the CV-TMLE estimate of the static rule in which everyone receives CBT (difference: -0.35\%, CI: [-6.40\%, 5.71\%]) and no one receives CBT (difference: -0.18\%, CI: [-7.06\%, 6.68\%]). Estimates and confidence intervals of these CV-TMLE estimates are illustrated in Figure \ref{eydopt}. Thus, there is insufficient evidence to conclude that assigning CBT using the ODTR SuperLearner is better than assigning CBT in a non-individualized way.

\section{Conclusions}

The aim of this paper was to illustrate the performance of different estimators that can be used to evaluate dynamic treatment rules, and in particular, the ODTR. At sample size 1,000, we saw a small price and many benefits to using CV-TMLE in order to estimate the following parameters: 1) the true value of a given \textit{a priori} known rule; 2) the true value of the true, unknown ODTR; and, 3) the true value of an estimated ODTR (a data-adaptive parameter). Of note, we see similar results when there is dependence between covariates in the DGP, as shown in the Appendix. In addition, we illustrated how to implement the CV-TMLE estimator to evaluate the ODTR using the ``Interventions" data as an applied example.

When evaluating estimators' performance for the value of a known rule, CV-TMLE performed well, irrespective of how data-adaptive the algorithms used for estimating nuisance parameters were. Although no estimator under an estimated ODTR yielded satisfactory performance for a target parameter corresponding to the true value of the true ODTR, when nuisance parameters and ODTRs were estimated using the most data-adaptive algorithms, CV-TMLE performed the best among the candidate estimators, while non-cross-validated estimators yielded overly optimistic and highly variable results. Finally, no estimator except CV-TMLE performed well when estimating a data-adaptive parameter -- a parameter that may be of interest if: 1) one believes one's estimate of the ODTR will not converge appropriately to its truth (as was the case for these estimators of the ODTR under the current DGP); and 2) one cares more about the performance of the estimated ODTR that is generated by the sample at hand (as opposed to the true, but unknown, ODTR). That said, the superior performance of CV-TMLE does come at the cost of estimating a distinct, data-adaptive parameter that depends not only on the sample at hand, but also the sample split. %MP: not sure I understand where to integrate this comment: ``outcome regression (either misspec parametric model or data-adaptive) not enough. if going to estimate the outcome regression, need to use a DR estimator. these are worked fine. in tis particular sim, no efficiency gains, but not expected to hold up generally (eg if have a covariate more predictive of the outcome)."

Future directions for simulations should evaluate results under varying sample sizes. In particular, for small sample sizes and thus less support in the data, it may be that case that we pay a price in performance by sample splitting. Additionally, future work could extend these simulations to the multiple time-point setting to evaluate the \emph{sequential} ODTR that could be generated from, for example, a SMART design \citep{lei2012smart, kosorok2015adaptive, almirall2014introduction} instead of an single time-point experiment. 

As an illustration of how to apply the ODTR SuperLearner to real data, we estimated the ODTR using the ``Interventions" Study to determine which types of criminal justice-involved adults with mental illness should be assigned CBT versus TAU, to yield the highest probability of no re-arrest. In our applied example using the ``Interventions" data, preliminary results suggest the probability of recidivism if treatment were assigned using the ODTR algorithm (i.e., in an individualized way) is not significantly different from probability of recidivism if all had been assigned treatment or no treatment (i.e., in a non-individualized way). This may indicate an absence of strong heterogeneous treatment effects by the measured variables, or it may reflect limitations in power to detect such effects due to preliminary sample sizes. In future work, we will apply the ODTR SuperLearner and evaluate it on the full sample size ($n = 720$).

This work contributes to statistical methods for understanding treatment effect heterogeneity, and in particular, how much improvement we might make in outcomes if interventions are assigned according to an ODTR. It is of great practical relevance to study estimators of these parameters, which allow us to determine the benefit of assigning treatment in a more individualized way compared to, for example, simply giving all participants treatment.

\section{Acknowledgments}

Research reported in this publication was supported by the National Institute Of
Allergy And Infectious Diseases of the National Institutes of Health under Award
Number R01AI074345 and F31AI140962. The content is solely the responsibility
of the authors and does not necessarily represent the official views of the National
Institutes of Health.

%% BibTeX support
\bibliographystyle{unsrt}
\bibliography{sample}

\begin{thebibliography}{10}

\bibitem{khoury2016precision}
Muin~J Khoury, Michael~F Iademarco, and William~T Riley.
\newblock Precision public health for the era of precision medicine.
\newblock {\em American journal of preventive medicine}, 50(3):398, 2016.

\bibitem{laber2017dynamic}
Eric Laber and Marie Davidian.
\newblock Dynamic treatment regimes, past, present, and future: A conversation
  with experts.
\newblock {\em Statistical methods in medical research}, 26(4):1605--1610,
  2017.

\bibitem{bembomvdL2007}
Oliver Bembom and Mark~J. van~der Laan.
\newblock A practical illustration of the importance of realistic
  individualized treatment rules in causal inference.
\newblock {\em Electron J Stat}, 1:574--596, 2007.

\bibitem{vdLpetersen2007}
Mark~J. van~der Laan and Maya~L. Petersen.
\newblock Causal effect models for realistic individualized treatment and
  intention to treat rules.
\newblock {\em Int J Biostat}, 3(1):Article 3, 2007.

\bibitem{robins1986new}
James Robins.
\newblock A new approach to causal inference in mortality studies with a
  sustained exposure period—application to control of the healthy worker
  survivor effect.
\newblock {\em Mathematical modelling}, 7(9-12):1393--1512, 1986.

\bibitem{chakraborty2013}
Bibhas Chakraborty, Eric Laber, and Yingqi Zhao.
\newblock Inference for optimal dynamic treatment regimes using an adaptive
  m-out-of-n bootstrap scheme.
\newblock {\em Biometrics}, 69(3):714--23, 2013.

\bibitem{Chakraborty2014}
Bibhas Chakraborty and Susan~A Murphy.
\newblock Dynamic treatment regimes.
\newblock {\em Annual review of statistics and its application}, 1:447--464,
  2014.

\bibitem{murphy2003}
Susan~A Murphy.
\newblock Optimal dynamic treatment regimes.
\newblock {\em Journal of the Royal Statistical Society: Series B (Statistical
  Methodology)}, 65(2):331--355, 2003.

\bibitem{robins2004}
James~M Robins.
\newblock Optimal structural nested models for optimal sequential decisions.
\newblock In {\em Proceedings of the second seattle Symposium in
  Biostatistics}, pages 189--326. Springer, 2004.

\bibitem{moodie2007}
E.~E. Moodie, T.~S. Richardson, and D.~A. Stephens.
\newblock Demystifying optimal dynamic treatment regimes.
\newblock {\em Biometrics}, 63(2):447--55, 2007.

\bibitem{kosorok2019}
M.~R. Kosorok and E.~B. Laber.
\newblock Precision medicine.
\newblock {\em Annu Rev Stat Appl}, 6:263--286, 2019.

\bibitem{kosorok2015adaptive}
Michael~R Kosorok and Erica~EM Moodie.
\newblock {\em Adaptive Treatment Strategies in Practice: Planning Trials and
  Analyzing Data for Personalized Medicine}, volume~21.
\newblock SIAM, 2015.

\bibitem{tsiatis2019dynamic}
Anastasios~A Tsiatis.
\newblock {\em Dynamic Treatment Regimes: Statistical Methods for Precision
  Medicine}.
\newblock CRC Press, 2019.

\bibitem{van2007super}
Mark~J van~der Laan, Eric~C Polley, and Alan~E Hubbard.
\newblock Super learner.
\newblock {\em Statistical applications in genetics and molecular biology},
  6(1), 2007.

\bibitem{luedtkeSLODTR}
Alexander~R. Luedtke and Mark~J. van~der Laan.
\newblock Super-learning of an optimal dynamic treatment rule.
\newblock {\em Int J Biostat}, 12(1):305--32, 2016.

\bibitem{coyle2017computational}
Jeremy~Robert Coyle.
\newblock {\em Computational Considerations for Targeted Learning}.
\newblock PhD thesis, UC Berkeley, 2017.

\bibitem{montoyaplaceholder}
Lina Montoya, Mark van~der Laan, Alexander Luedtke, Jennifer Skeem, Jeremy
  Coyle, and Maya Petersen.
\newblock The optimal dynamic treatment rule superlearner: Considerations,
  performance, and application.
\newblock {\em arXiv preprint arXiv:2101.12326}, 2021.

\bibitem{hernanrobins2006}
Miguel~A. Hernan and James~M. Robins.
\newblock Estimating causal effects from epidemiological data.
\newblock {\em J Epidemiol Community Health}, 60(7):578--86, 2006.

\bibitem{rosenbaum1983central}
Paul~R Rosenbaum and Donald~B Rubin.
\newblock The central role of the propensity score in observational studies for
  causal effects.
\newblock {\em Biometrika}, 70(1):41--55, 1983.

\bibitem{robins1994estimation}
James~M Robins, Andrea Rotnitzky, and Lue~Ping Zhao.
\newblock Estimation of regression coefficients when some regressors are not
  always observed.
\newblock {\em Journal of the American statistical Association},
  89(427):846--866, 1994.

\bibitem{scharfstein1999theory}
Daniel~O Scharfstein, Andrea Rotnitzky, and James~M Robins.
\newblock Theory and methods-rejoinder-adjusting for nonignorable drop-out
  using semiparametric nonresponse models.
\newblock {\em Journal of the American Statistical Association},
  94(448):1135--1146, 1999.

\bibitem{robins2000robust}
James~M Robins.
\newblock Robust estimation in sequentially ignorable missing data and causal
  inference models.
\newblock In {\em Proceedings of the American Statistical Association}, volume
  1999, pages 6--10. Indianapolis, IN, 2000.

\bibitem{rosenblum2010targeted}
Michael Rosenblum and Mark~J van~der Laan.
\newblock Targeted maximum likelihood estimation of the parameter of a marginal
  structural model.
\newblock {\em The international journal of biostatistics}, 6(2), 2010.

\bibitem{TLBBD}
Mark~J van~der Laan and Sherri Rose.
\newblock {\em Targeted learning: causal inference for observational and
  experimental data}.
\newblock Springer Science \& Business Media, 2011.

\bibitem{van2015targeted}
Mark~J van~der Laan and Alexander~R Luedtke.
\newblock Targeted learning of the mean outcome under an optimal dynamic
  treatment rule.
\newblock {\em Journal of causal inference}, 3(1):61--95, 2015.

\bibitem{zheng2010asymptotic}
Wenjing Zheng and Mark~J van~der Laan.
\newblock Asymptotic theory for cross-validated targeted maximum likelihood
  estimation.
\newblock 2010.

\bibitem{van2018targeted}
Mark~J van~der Laan and Sherri Rose.
\newblock {\em Targeted Learning in Data Science}.
\newblock Springer, 2018.

\bibitem{laberchapter2017}
Eric Laber and Min Qian.
\newblock Evaluating personalized treatment regimes.
\newblock In Constantine Gatsonis and Sally~C. Morton, editors, {\em Methods in
  Comparative Effectiveness Research}, book chapter~15, pages 483--497. CRC
  Press LLC : Chapman and Hall/CRC, Boca Raton, FL, 2017.

\bibitem{chakraborty2010}
Bibhas Chakraborty, Susan Murphy, and Victor Strecher.
\newblock Inference for non-regular parameters in optimal dynamic treatment
  regimes.
\newblock {\em Statistical methods in medical research}, 19(3):317--343, 2010.

\bibitem{sies2019}
A.~Sies and I.~Van~Mechelen.
\newblock Estimating the quality of optimal treatment regimes.
\newblock {\em Stat Med}, 2019.

\bibitem{hubbard2016}
A.~E. Hubbard, S.~Kherad-Pajouh, and M.~J. van~der Laan.
\newblock Statistical inference for data adaptive target parameters.
\newblock {\em Int J Biostat}, 12(1):3--19, 2016.

\bibitem{skeem2011correctional}
Jennifer~L Skeem, Sarah Manchak, and Jillian~K Peterson.
\newblock Correctional policy for offenders with mental illness: Creating a new
  paradigm for recidivism reduction.
\newblock {\em Law and human behavior}, 35(2):110--126, 2011.

\bibitem{skeem2014offenders}
Jennifer~L Skeem, Eliza Winter, Patrick~J Kennealy, Jennifer~Eno Louden, and
  Joseph~R Tatar~II.
\newblock Offenders with mental illness have criminogenic needs, too: Toward
  recidivism reduction.
\newblock {\em Law and human behavior}, 38(3):212, 2014.

\bibitem{petersen2014causal}
Maya~L Petersen and Mark~J van~der Laan.
\newblock Causal models and learning from data: integrating causal modeling and
  statistical estimation.
\newblock {\em Epidemiology (Cambridge, Mass.)}, 25(3):418, 2014.

\bibitem{petersen2012diagnosing}
Maya~L Petersen, Kristin~E Porter, Susan Gruber, Yue Wang, and Mark~J van~der
  Laan.
\newblock Diagnosing and responding to violations in the positivity assumption.
\newblock {\em Statistical methods in medical research}, 21(1):31--54, 2012.

\bibitem{luedtke2016statistical}
Alexander~R Luedtke and Mark~J van~der Laan.
\newblock Statistical inference for the mean outcome under a possibly
  non-unique optimal treatment strategy.
\newblock {\em Annals of statistics}, 44(2):713, 2016.

\bibitem{gruber2010targeted}
Susan Gruber and Mark~J van~der Laan.
\newblock A targeted maximum likelihood estimator of a causal effect on a
  bounded continuous outcome.
\newblock {\em The International Journal of Biostatistics}, 6(1), 2010.

\bibitem{van2003unified}
Mark~J van~der Laan, MJ~Laan, and James~M Robins.
\newblock {\em Unified methods for censored longitudinal data and causality}.
\newblock Springer Science \& Business Media, 2003.

\bibitem{van2000asymptotic}
Aad~W Van~der Vaart.
\newblock {\em Asymptotic statistics}, volume~3.
\newblock Cambridge university press, 2000.

\bibitem{bickel1993efficient}
Peter~J Bickel, Chris~AJ Klaassen, Peter~J Bickel, Ya’acov Ritov, J~Klaassen,
  Jon~A Wellner, and YA'Acov Ritov.
\newblock {\em Efficient and adaptive estimation for semiparametric models},
  volume~4.
\newblock Johns Hopkins University Press Baltimore, 1993.

\bibitem{R}
R~Core Team.
\newblock {\em R: A Language and Environment for Statistical Computing}.
\newblock R Foundation for Statistical Computing, Vienna, Austria, 2018.

\bibitem{tlverse}
Jeremy Coyle.
\newblock {\em tlverse: Umbrella Package for Targeted Learning in R}, 2021.
\newblock R package version 0.0.1.

\bibitem{friedman1991multivariate}
Jerome~H Friedman et~al.
\newblock Multivariate adaptive regression splines.
\newblock {\em The annals of statistics}, 19(1):1--67, 1991.

\bibitem{ripley1996pattern}
Brian~D Ripley and NL~Hjort.
\newblock {\em Pattern recognition and neural networks}.
\newblock Cambridge university press, 1996.

\bibitem{chang2011libsvm}
Chih-Chung Chang and Chih-Jen Lin.
\newblock Libsvm: A library for support vector machines.
\newblock {\em ACM transactions on intelligent systems and technology (TIST)},
  2(3):27, 2011.

\bibitem{breiman2017classification}
Leo Breiman.
\newblock {\em Classification and regression trees}.
\newblock Routledge, 2017.

\bibitem{SLpackage}
Eric Polley, Erin LeDell, Chris Kennedy, and Mark van~der Laan.
\newblock {\em SuperLearner: Super Learner Prediction}, 2018.
\newblock R package version 2.0-24.

\bibitem{breiman2001random}
Leo Breiman.
\newblock Random forests.
\newblock {\em Machine learning}, 45(1):5--32, 2001.

\bibitem{rosenblum2009using}
Michael Rosenblum and Mark~J van~der Laan.
\newblock Using regression models to analyze randomized trials: Asymptotically
  valid hypothesis tests despite incorrectly specified models.
\newblock {\em Biometrics}, 65(3):937--945, 2009.

\bibitem{lei2012smart}
Huitian Lei, Inbal Nahum-Shani, K~Lynch, David Oslin, and Susan~A Murphy.
\newblock A" smart" design for building individualized treatment sequences.
\newblock {\em Annual review of clinical psychology}, 8:21--48, 2012.

\bibitem{almirall2014introduction}
Daniel Almirall, Inbal Nahum-Shani, Nancy~E Sherwood, and Susan~A Murphy.
\newblock Introduction to smart designs for the development of adaptive
  interventions: with application to weight loss research.
\newblock {\em Translational behavioral medicine}, 4(3):260--274, 2014.

\end{thebibliography}

\begin{table}
\begin{tabular}{l|l|l}
\hline
  & TAU ($A=0$) & CBT ($A=1$) \\
\hline
$n$ & 211 & 230    \\
\hline
\textbf{No re-arrest} ($Y=1$) (\%) & 128 (60.7) & 143 (62.2) \\
\hline
\textbf{Site} = San Francisco (\%) & 87 (41.2) & 104 (45.2) \\
\hline
\textbf{Gender} = Female (\%) & 38 (18.0) & 37 (16.1) \\
\hline
\textbf{Ethnicity} = Hispanic (\%) & 50 (23.7) & 42 (18.3) \\
\hline
\textbf{Age} (mean (SD)) & 38.08 (11.05) & 37.01 (11.22) \\
\hline
\textbf{CSI} (mean (SD)) & 32.35 (11.13) & 33.46 (11.27)  \\
\hline
\textbf{LSI} (mean (SD)) & 5.59 (1.33) & 5.50 (1.48)  \\
\hline
\textbf{SES} (mean (SD)) & 3.81 (1.89) & 3.81 (2.12)  \\
\hline
\textbf{Prior adult convictions} (\%) &  &   \\
\hline
\hspace{3mm} Zero to two times & 74 (35.1) & 93 (40.4) \\
\hline
\hspace{3mm} Three or more times & 134 (63.5) & 129 (56.1)  \\
\hline
\hspace{3mm} Missing & 3 (1.4) & 8 (3.5)   \\
\hline
\textbf{Most serious offense} (mean (SD)) & 5.29 (2.54) & 5.09 (2.52)  \\
\hline
\textbf{Motivation} (mean (SD)) & 3.22 (1.36) & 3.27 (1.37)  \\
\hline
\textbf{Substance use} (\%) &  &   \\
\hline
\hspace{3mm} 0 & 53 (25.1) & 76 (33.0) \\
\hline
\hspace{3mm} 1 & 47 (22.3) & 55 (23.9) \\
\hline
\hspace{3mm} 2 & 109 (51.7) & 98 (42.6) \\
\hline
\hspace{3mm} Missing & 2 (0.9) & 1 (0.4)   \\
\hline
\end{tabular}
\caption{Distribution of baseline covariates in the ``Interventions" data set, stratified by randomized treatment assignment (TAU denotes Treatment as Usual, CBT denotes Cognitive Behavioral Therapy).}
\label{table_descriptives}
\end{table}

\begin{figure}[h]
    \centering
    \includegraphics[scale = .35]{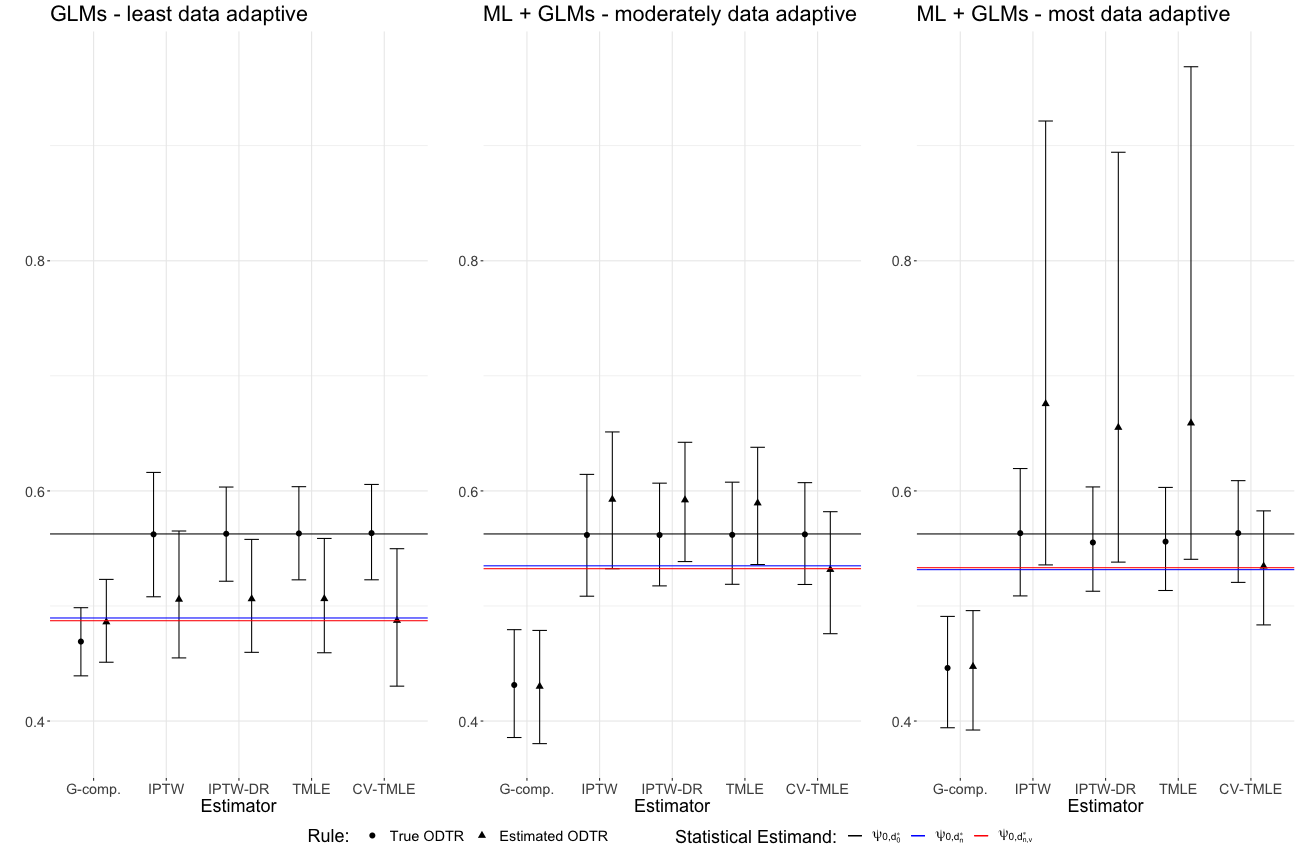}
    \caption{Performance of candidate estimators of the value of a given rule, evaluated for 3 SuperLearner library configurations with increasing (left to right) levels of data-adaptivity used for estimating the true outcome regression $Q_0$ and/or the true optimal rule $d^*_0$ (``GLM - least data adaptive", ``ML + GLMs - moderately data adaptive", "ML + GLMs - most data adaptive"). The horizontal black line depicts the true mean outcome under the true ODTR $\psi_{0,d^*_0}$; the blue and red lines are true values of the data-adaptive parameters $\psi_{0,d^*_n}$ and $\psi_{0,d^*_{n,v}}$, respectively, averaged over each of the 1,000 simulated samples. Points with error bars show the distribution of the estimators across the 1,000 simulated samples (G-computation estimator, IPTW estimator, TMLE, and CV-TMLE); the points (circles and triangles) show the estimates averaged over the samples, and error bars show the $2.5^{th}$ and $97.5^{th}$ quantiles of the distribution of each estimator across the simulation repetitions. The circles depict the estimators under a known rule $\hat{\psi}_{d=d_0^*}$ and the triangles illustrate the estimators under an estimated rule, either $\hat{\psi}_{d^*_n}$ or $\hat{\psi}_{d_{n,v}^*}$ (for CV-TMLE).}
    \label{plotfig}
\end{figure}

% Please add the following required packages to your document preamble:
% \usepackage{multirow}
\begin{table}[]
\scalebox{0.9}{ 
\begin{tabular}{|l|l|l|l|l|l|}
\hline
Library & Estimator & Bias & Variance & MSE & Coverage \\ \hline
\multirow{5}{*}{GLMs - least data adaptive} & G-comp. & -0.0935 & 0.0002 & 0.0090 & - \\ \cline{2-6} 
 & IPTW & -0.0004 & 0.0008 & 0.0008 & 95.80\% \\ \cline{2-6} 
 & IPTW-DR & 0.0002 & 0.0004 & 0.0004 & 95.80\% \\ \cline{2-6} 
 & TMLE & 0.0004 & 0.0004 & 0.0004 & 95.80\% \\ \cline{2-6} 
 & CV-TMLE & 0.0007 & 0.0005 & 0.0005 & 95.30\% \\ \hline
\multirow{5}{*}{ML + GLMs - moderately data adaptive} & G-comp. & -0.1313 & 0.0006 & 0.0179 & - \\ \cline{2-6} 
 & IPTW & -0.0009 & 0.0007 & 0.0007 & 96.30\% \\ \cline{2-6} 
 & IPTW-DR & -0.0009 & 0.0005 & 0.0005 & 95.00\% \\ \cline{2-6} 
 & TMLE & -0.0008 & 0.0005 & 0.0005 & 94.80\% \\ \cline{2-6} 
 & CV-TMLE & -0.0004 & 0.0005 & 0.0005 & 94.90\% \\ \hline
\multirow{5}{*}{ML + GLMs - most data adaptive} & G-comp. & -0.1165 & 0.0006 & 0.0142 & - \\ \cline{2-6} 
 & IPTW & 0.0007 & 0.0008 & 0.0008 & 95.20\% \\ \cline{2-6} 
 & IPTW-DR & -0.0074 & 0.0005 & 0.0006 & 90.50\% \\ \cline{2-6} 
 & TMLE & -0.0067 & 0.0005 & 0.0006 & 91.00\% \\ \cline{2-6} 
 & CV-TMLE & 0.0007 & 0.0005 & 0.0005 & 94.70\% \\ \hline
\end{tabular}
}
\caption{Performance metrics (bias, variance, MSE, confidence interval coverage) of each estimator $\hat{\psi}_{d=d^*_0}$ (G-computation, Inverse Probability of Treatment Weighting [IPTW], Double-robust IPTW [IPTW-DR], Targeted Maximum Likelihood Estimation [TMLE], Cross-Validated TMLE [CV-TMLE]) of the true expected outcome under a given, known dynamic treatment rule ($\psi_{0,d^*_0}$), for each library configuration used to estimate the outcome regression ($Q_n$).}
\label{table1}
\end{table}

% Please add the following required packages to your document preamble:
% \usepackage{multirow}
\begin{table}[]
\scalebox{0.9}{ 
\begin{tabular}{|l|l|l|l|l|l|}
\hline
Library & Estimator & Bias & Variance & MSE & Coverage \\ \hline
\multirow{5}{*}{GLMs - least data adaptive} & G-comp. & -0.0765 & 0.0003 & 0.0062 & - \\ \cline{2-6} 
 & IPTW & -0.0569 & 0.0008 & 0.0041 & 45.70\% \\ \cline{2-6} 
 & IPTW-DR & -0.0565 & 0.0007 & 0.0038 & 29.80\% \\ \cline{2-6} 
 & TMLE & -0.0563 & 0.0007 & 0.0038 & 29.40\% \\ \cline{2-6} 
 & CV-TMLE & -0.0752 & 0.0009 & 0.0066 & 14.00\% \\ \hline
\multirow{5}{*}{ML + GLMs - moderately data adaptive} & G-comp. & -0.1325 & 0.0007 & 0.0182 & - \\ \cline{2-6} 
 & IPTW & 0.0300 & 0.0010 & 0.0019 & 79.50\% \\ \cline{2-6} 
 & IPTW-DR & 0.0295 & 0.0008 & 0.0016 & 69.90\% \\ \cline{2-6} 
 & TMLE & 0.0268 & 0.0007 & 0.0014 & 72.20\% \\ \cline{2-6} 
 & CV-TMLE & -0.0310 & 0.0007 & 0.0017 & 70.30\% \\ \hline
\multirow{5}{*}{ML + GLMs - most data adaptive} & G-comp. & -0.1152 & 0.0007 & 0.0140 & - \\ \cline{2-6} 
 & IPTW & 0.1131 & 0.0114 & 0.0242 & 38.30\% \\ \cline{2-6} 
 & IPTW-DR & 0.0925 & 0.0096 & 0.0181 & 38.90\% \\ \cline{2-6} 
 & TMLE & 0.0963 & 0.0118 & 0.0211 & 38.50\% \\ \cline{2-6} 
 & CV-TMLE & -0.0281 & 0.0007 & 0.0015 & 71.90\% \\ \hline
\end{tabular}
}
\caption{Performance metrics (bias, variance, MSE, confidence interval coverage) of each estimator $\hat{\psi}_{d^*_n}$ (G-computation, Inverse Probability of Treatment Weighting [IPTW], Double-robust IPTW [IPTW-DR], Targeted Maximum Likelihood Estimation [TMLE]) or $\hat{\psi}_{d^*_{n,v}}$ (Cross-validated-TMLE [CV-TMLE]) of the true expected outcome under the true optimal dynamic treatment rule (ODTR; $\psi_{0,d^*_0}$), for each library configuration used to estimate the outcome regression ($Q_n$) and the ODTR ($d^*_n$).}
\label{table2}
\end{table}

% Please add the following required packages to your document preamble:
% \usepackage{multirow}
\begin{table}[]
\scalebox{0.9}{ 
\begin{tabular}{|l|l|l|l|l|l|}
\hline
Library & Estimator & Bias & Variance & MSE & Coverage \\ \hline
\multirow{5}{*}{GLMs - least data adaptive} & G-comp. & -0.0035 & 0.0003 & 0.0004 & - \\ \cline{2-6} 
 & IPTW & 0.0162 & 0.0008 & 0.0011 & 94.90\% \\ \cline{2-6} 
 & IPTW-DR & 0.0166 & 0.0007 & 0.0009 & 90.60\% \\ \cline{2-6} 
 & TMLE & 0.0167 & 0.0007 & 0.0009 & 90.50\% \\ \cline{2-6} 
 & CV-TMLE & 0.0002 & 0.0009 & 0.0009 & 93.90\% \\ \hline
\multirow{5}{*}{ML + GLMs - moderately data adaptive} & G-comp. & -0.1048 & 0.0007 & 0.0117 & - \\ \cline{2-6} 
 & IPTW & 0.0577 & 0.001 & 0.0043 & 48.00\% \\ \cline{2-6} 
 & IPTW-DR & 0.0572 & 0.0008 & 0.0041 & 33.00\% \\ \cline{2-6} 
 & TMLE & 0.0545 & 0.0007 & 0.0037 & 33.90\% \\ \cline{2-6} 
 & CV-TMLE & -0.0008 & 0.0007 & 0.0007 & 93.90\% \\ \hline
\multirow{5}{*}{ML + GLMs - most data adaptive} & G-comp. & -0.0842 & 0.0007 & 0.0078 & - \\ \cline{2-6} 
 & IPTW & 0.1442 & 0.0114 & 0.0322 & 25.00\% \\ \cline{2-6} 
 & IPTW-DR & 0.1236 & 0.0096 & 0.0248 & 22.60\% \\ \cline{2-6} 
 & TMLE & 0.1274 & 0.0118 & 0.0280 & 22.50\% \\ \cline{2-6} 
 & CV-TMLE & 0.0013 & 0.0007 & 0.0007 & 93.90\% \\ \hline
\end{tabular}
}
\caption{Performance metrics (bias, variance, MSE, confidence interval coverage) of each estimator $\hat{\psi}_{d^*_n}$ (G-computation, Inverse Probability of Treatment Weighting [IPTW], Double-robust IPTW [IPTW-DR], Targeted Maximum Likelihood Estimation [TMLE]) of the true expected outcome under the sample-specific estimate of the optimal dynamic treatment rule (ODTR; $\psi_{0,d^*_n}$) or $\hat{\psi}_{d^*_{n,v}}$ (Cross-validated-TMLE [CV-TMLE]) of the true expected outcome under the sample-split-specific estimate of the ODTR ($\psi_{0,d^*_{n,v}}$), for each library configuration used to estimate the outcome regression ($Q_n$) and the ODTR ($d^*_n$).}
\label{table3}
\end{table}

\section{Appendix}

\subsection{Notation Table}

{\tabulinesep=0.8mm
\begin{longtabu}[]{ll}
{{\ul \textbf{Notation}}} & {\ul \textbf{Definition}} \\
\textbf{Observed Data Random Variables} &  \\
$W$ & Vector of covariates \\
$A$ & Treatment or exposure \\
$Y$ & Outcome \\
\textbf{Counterfactual outcomes} &  \\
$Y_1$ & Counterfactual outcome under treatment \\
$Y_0$ & Counterfactual outcome under control \\
$Y_d$ & Counterfactual outcome under dynamic treatment rule \\
\textbf{Models and distributions} &  \\
$\mathcal{M}^F$ & Causal model \\
$\mathcal{M}$ & Statistical model \\
$P_{U,X}$ & \begin{tabular}[c]{@{}l@{}}Probability distribution for full data $X$ (including \\ counterfactual outcomes) and unobserved \\ variables $U$. $P_{U,X} \in \mathcal{M}^F$.\end{tabular} \\
$P_0$ & \begin{tabular}[c]{@{}l@{}}True probability distribution for observed data $O$. \\ $P_{0} \in \mathcal{M}.$\end{tabular} \\
$P_n$ & \begin{tabular}[c]{@{}l@{}}Empirical probability distribution that places \\ weight $\frac{1}{n}$ on each observation\end{tabular} \\
$P_{n,v}$ & \begin{tabular}[c]{@{}l@{}}Empirical probability distribution of the validation \\ sample\end{tabular} \\
$P_{n,-v}$ & \begin{tabular}[c]{@{}l@{}}Empirical probability distribution of the training \\ sample\end{tabular} \\
\textbf{Causal parameters} &  \\
$\Psi^F_d(P_{U,X}) = \mathbb{E}_{P_{U,X}}[Y_d]$ & \begin{tabular}[c]{@{}l@{}}Expected counterfactual outcome (value) had everyone \\ received a given dynamic treatment rule $d$\end{tabular} \\
$\mathbb{E}_{P_{U,X}}[Y_1-Y_0|W]$ & Conditional average treatment effect (CATE) \\
$\Psi^F_{d_0^*}(P_{U,X}) = \mathbb{E}_{P_{U,X}} [Y_{d_0^*}]$ & \begin{tabular}[c]{@{}l@{}}Expected counterfactual outcome (value) had everyone \\ received the ODTR $d_0^*$\end{tabular} \\
$\Psi^F_{d^*_n}(P_{U,X}) = \mathbb{E}_{P_{U,X}} [Y_{d^*_n}]$ & \begin{tabular}[c]{@{}l@{}}Expected counterfactual outcome (value) had everyone \\ received the sample-specific estimate of the ODTR $d_n^*$\end{tabular} \\
$\Psi^F_{d^*_{n,v}}(P_{U,X}) = \frac{1}{V}\sum_{v=1}^V \mathbb{E}_{P_{U,X}} [Y_{d^*_{n,v}}]$ & \begin{tabular}[c]{@{}l@{}}Average of the expected validation set counterfactual \\ outcomes (values) under the training set-specific \\ estimates of the ODTR $d_{n,v}^*$\end{tabular} \\
\textbf{Statistical parameters} &  \\
$g_0(A \vert W)$ & \begin{tabular}[c]{@{}l@{}}True probability of treatment given covariates \\ (i.e., treatment mechanism; $P_0(A\vert W)$)\end{tabular} \\
$Q_0(A,W)$ & \begin{tabular}[c]{@{}l@{}}True conditional mean outcome given treatment \\ and covariates (i.e., outcome regression; $\mathbb{E}[Y\vert A,W]$)\end{tabular} \\
$d_0^*(W)$ & True optimal dynamic treatment rule (ODTR) \\
$B_0(W)$ & Blip function (identification result of the CATE) \\
$\psi_{0,d} = \Psi_{d}(P_0) =\mathbb{E}_0[Q_{0}(d(W),W)]$ & True value of a given dynamic treatment rule $d$ \\
$\psi_{0,d^*_0} = \Psi_{d^*_0}(P_0) = \mathbb{E}_0[Q_{0}(d^*_0(W),W)]$ & True value of the ODTR \\
$\psi_{0,d^*_n} \equiv \Psi_{d^*_n}(P_0) = \mathbb{E}_0[Q_{0}(d^*_n(W),W)]$ & True value of the sample-specific estimate of the ODTR \\
\begin{tabular}[c]{@{}l@{}}$\psi_{0,d^*_{n,v}} = \Psi_{d^*_{n,v}}(P_0) $\\ $=  \frac{1}{V}\sum_{v=1}^V\mathbb{E}_0[Q_{0}(d^*_{n,v}(W),W)]$\end{tabular} & \begin{tabular}[c]{@{}l@{}}Average of the true values of the training set-specific \\ estimates of the ODTR\end{tabular} \\
\textbf{Estimates} &  \\
$g_n(A \vert W)$ & Estimate of the true treatment mechanism \\
$Q_n(A,W)$ & Estimate of the outcome regression \\
$D_n(Q_n,g_n)(O)$ & \begin{tabular}[c]{@{}l@{}}Estimated pseudo-outcome. Regress this on \\ $W$ to get estimate of blip.\end{tabular} \\
$d_n^*(W)$ & ODTR estimated on the entire sample \\
$d_{n,v}^*(W)$ & Training sample-specific estimate of the ODTR \\
$d_{n,j}(W)$ & Estimate of ODTR under $j^{th}$ candidate algorithm \\
$d_{n,\alpha}(W)$ & \begin{tabular}[c]{@{}l@{}}A convex combination of candidate ODTR estimates, \\ with algorithm-specific weights given by $\alpha$\end{tabular} \\
$d_{n,B}^*(W), d_{n,d}^*$ & \begin{tabular}[c]{@{}l@{}}ODTR estimated on the entire sample  \\ using blip and direct estimation approaches, respectively\end{tabular} \\
$B_n(W)$ & Estimate of the blip \\
$B_{n,j}(W)$ & Estimate of blip under $j^{th}$ candidate algorithm \\
$B_{n,\alpha}(W)$ & \begin{tabular}[c]{@{}l@{}}A convex combination of candidate blip estimates,\\ with algorithm-specific weights given by $\alpha$\end{tabular} \\
$\hat{\psi}_d = \hat{\Psi}_d(P_n)$ & \begin{tabular}[c]{@{}l@{}}Estimate of the value of a given dynamic treatment \\ rule $d$ (estimator is either IPTW, IPTW-DR, \\ TMLE, or CV-TMLE)\end{tabular} \\
$\hat{\psi}_{d_n^*} = \hat{\Psi}_{d_n^*}(P_n)$ & \begin{tabular}[c]{@{}l@{}}Estimate of the value of sample-specific estimate \\ of the ODTR (estimator is either IPTW, \\ IPTW-DR, or TMLE)\end{tabular} \\
$\hat{\psi}_{d_{n,v}^*} = \hat{\Psi}_{d_{n,v}^*}(P_n)$ & \begin{tabular}[c]{@{}l@{}}Estimate of the value of training-set specific \\ estimate of the ODTR (estimator is CV-TMLE)\end{tabular} \\
\textbf{Risk functions} &  \\
$R_{MSE}$ & \begin{tabular}[c]{@{}l@{}}Risk for ODTR SuperLearner targeting the blip \\ function using mean-squared error\end{tabular} \\
$R_{E[Y_d]}$ & \begin{tabular}[c]{@{}l@{}}Risk for ODTR SuperLearner targeting the \\ expected rule-specific outcome\end{tabular} \\
\textbf{Other} &  \\
$d(W)$ & \begin{tabular}[c]{@{}l@{}}A given dynamic treatment rule. A function that takes \\ as input covariates $W$ (or function of $W$) and \\ outputs a treatment decision.\end{tabular} \\
$\mathcal{D}$ & Set of all dynamic treatment rules \\
$\alpha = \{\alpha_1,...,\alpha_J\}$ & \begin{tabular}[c]{@{}l@{}}Weight vector for convex combination of \\ $J$ candidate algorithms\end{tabular} \\
$\alpha_n$ & \begin{tabular}[c]{@{}l@{}}``Best" weighting of the algorithms, i.e., convex \\ combination $\alpha$ that yields the smallest cross-validated \\ empirical risk\end{tabular} \\
$\widehat{IC}(O)$ & \begin{tabular}[c]{@{}l@{}}Working influence curve of an estimator, a function \\ of the observed data $O$\end{tabular}
\end{longtabu}}

\subsection{Simulation Extension to DGP with Dependent Covariates}

We include extra simulations illustrating a DGP scenario where covariates are dependent, as follows:

\begin{align*}
W_1 &\sim Normal(\mu=0,\sigma^2=1) \\
W_2,W_3,W_4 &\sim \mathcal{N}\left(\mathbold{\mu} = \begin{bmatrix}
0\\
0
\end{bmatrix}, \mathbold{\Sigma} = \begin{bmatrix}
1.0 & 0.3 & 0.7\\
0.3 & 1.0 & 0.8\\
0.7 & 0.8 & 1.0
\end{bmatrix}\right) \\
A &\sim Bernoulli(p=0.5) \\
Y \sim & Bernoulli(p=Q_0(A,W)),\\
\text{where } Q_0(A,W) =& 0.5\textrm{expit} (1-W_1^2  + 3W_2  + 5W_3^2 A - 4.45A)+ \\
& 0.5\textrm{expit} (-0.5- W_3  + 2W_1 W_2  + 3|W_2|A - 1.5A),
\end{align*}
then the true blip function is:
\begin{align*}
    B_0 (W)= & 0.5[\textrm{expit} (1-W_1^2  + 3W_2  + 5W_3^2  - 4.45)+\textrm{expit} (-0.5- W_3  + 2W_1 W_2  + 3|W_2|  - 1.5)\\
& - \textrm{expit} (1-W_1^2  + 3W_2 )-\textrm{expit} (-0.5- W_3  + 2W_1 W_2 )].
\end{align*}

The true expected outcome under the true ODTR $\Psi^F_{d_0^*}(P_{U,X}) \approx 0.5595$ and the true optimal proportion treated $\mathbb{E}_{P_{U,X} } [d_0^* ] \approx 54.0\%$. The mean outcome had everyone and no one been treated are, respectively, $\mathbb{E}_{P_{U,X}} [Y_1 ] \approx 0.5152$ and $\mathbb{E}_{P_{U,X}} [Y_0 ] \approx 0.5000$.

% Please add the following required packages to your document preamble:
% \usepackage{multirow}
\begin{table}[]
\begin{tabular}{|l|l|l|l|l|l|}
\hline
Library & Estimator & Bias & Variance & MSE & Coverage \\ \hline
\multirow{5}{*}{GLMs - least data adaptive} & G-comp. & -0.0917 & 0.0002 & 0.0086 & - \\ \cline{2-6} 
 & IPTW & 0.0032 & 0.0007 & 0.0007 & 95.40\% \\ \cline{2-6} 
 & IPTW-DR & 0.0029 & 0.0004 & 0.0004 & 97.10\% \\ \cline{2-6} 
 & TMLE & 0.0030 & 0.0004 & 0.0004 & 96.80\% \\ \cline{2-6} 
 & CV-TMLE & 0.0031 & 0.0004 & 0.0004 & 96.20\% \\ \hline
\multirow{5}{*}{ML + GLMs - moderately data adaptive} & G-comp. & -0.0954 & 0.0012 & 0.0103 & - \\ \cline{2-6} 
 & IPTW & 0.0037 & 0.0008 & 0.0008 & 94.80\% \\ \cline{2-6} 
 & IPTW-DR & 0.0035 & 0.0005 & 0.0005 & 95.20\% \\ \cline{2-6} 
 & TMLE & 0.0034 & 0.0005 & 0.0005 & 95.90\% \\ \cline{2-6} 
 & CV-TMLE & 0.0037 & 0.0005 & 0.0005 & 95.20\% \\ \hline
\multirow{5}{*}{ML + GLMs - most data adaptive} & G-comp. & -0.0875 & 0.0008 & 0.0085 & - \\ \cline{2-6} 
 & IPTW & 0.0031 & 0.0008 & 0.0008 & 94.30\% \\ \cline{2-6} 
 & IPTW-DR & -0.0041 & 0.0005 & 0.0005 & 92.70\% \\ \cline{2-6} 
 & TMLE & -0.0033 & 0.0005 & 0.0005 & 92.90\% \\ \cline{2-6} 
 & CV-TMLE & 0.0034 & 0.0005 & 0.0005 & 94.20\% \\ \hline
\end{tabular}
\caption{Performance metrics of each estimator $\hat{\psi}_{d=d^*_0}$ for $\psi_{0,d^*_0}$, for each library configuration of $Q_n$.}
\end{table}

% Please add the following required packages to your document preamble:
% \usepackage{multirow}
\begin{table}[]
\begin{tabular}{|l|l|l|l|l|l|}
\hline
Library & Estimator & Bias & Variance & MSE & Coverage \\ \hline
\multirow{5}{*}{GLMs - least data adaptive} & G-comp. & -0.0720 & 0.0004 & 0.0055 & - \\ \cline{2-6} 
 & IPTW & -0.0527 & 0.0009 & 0.0036 & 49.10\% \\ \cline{2-6} 
 & IPTW-DR & -0.0531 & 0.0007 & 0.0035 & 35.20\% \\ \cline{2-6} 
 & TMLE & -0.0531 & 0.0007 & 0.0035 & 35.20\% \\ \cline{2-6} 
 & CV-TMLE & -0.0705 & 0.0010 & 0.0059 & 19.30\% \\ \hline
\multirow{5}{*}{ML + GLMs - moderately data adaptive} & G-comp. & -0.0953 & 0.0013 & 0.0104 & - \\ \cline{2-6} 
 & IPTW & 0.0227 & 0.0009 & 0.0014 & 86.70\% \\ \cline{2-6} 
 & IPTW-DR & 0.0263 & 0.0007 & 0.0013 & 74.30\% \\ \cline{2-6} 
 & TMLE & 0.0259 & 0.0006 & 0.0013 & 74.30\% \\ \cline{2-6} 
 & CV-TMLE & -0.0233 & 0.0007 & 0.0012 & 77.20\% \\ \hline
\multirow{5}{*}{ML + GLMs - most data adaptive} & G-comp. & -0.0848 & 0.0010 & 0.0082 & - \\ \cline{2-6} 
 & IPTW & 0.1000 & 0.0095 & 0.0195 & 43.00\% \\ \cline{2-6} 
 & IPTW-DR & 0.0860 & 0.0080 & 0.0154 & 38.90\% \\ \cline{2-6} 
 & TMLE & 0.0938 & 0.0106 & 0.0194 & 37.80\% \\ \cline{2-6} 
 & CV-TMLE & -0.0221 & 0.0006 & 0.0011 & 78.70\% \\ \hline
\end{tabular}
\caption{Performance metrics of each estimator $\hat{\psi}_{d^*_n}$ (G-computation, IPTW, IPTW-DR, TMLE) or $\hat{\psi}_{d^*_{n,v}}$ (CV-TMLE) for $\psi_{0,d^*_0}$, for each library configuration of $Q_n$ and $d^*_n$.}
\end{table}

% Please add the following required packages to your document preamble:
% \usepackage{multirow}
\begin{table}[]
\begin{tabular}{|l|l|l|l|l|l|}
\hline
Library & Estimator & Bias & Variance & MSE & Coverage \\ \hline
\multirow{5}{*}{GLMs - least data adaptive} & G-comp. & -0.0029 & 0.0004 & 0.0004 & - \\ \cline{2-6} 
 & IPTW & 0.0164 & 0.0009 & 0.0011 & 96.10\% \\ \cline{2-6} 
 & IPTW-DR & 0.0160 & 0.0007 & 0.0010 & 91.20\% \\ \cline{2-6} 
 & TMLE & 0.0160 & 0.0007 & 0.0010 & 91.00\% \\ \cline{2-6} 
 & CV-TMLE & 0.0009 & 0.0010 & 0.0010 & 92.70\% \\ \hline
\multirow{5}{*}{ML + GLMs - moderately data adaptive} & G-comp. & -0.0745 & 0.0013 & 0.0069 & - \\ \cline{2-6} 
 & IPTW & 0.0435 & 0.0009 & 0.0028 & 69.30\% \\ \cline{2-6} 
 & IPTW-DR & 0.0472 & 0.0007 & 0.0029 & 45.30\% \\ \cline{2-6} 
 & TMLE & 0.0467 & 0.0006 & 0.0028 & 43.80\% \\ \cline{2-6} 
 & CV-TMLE & -0.0002 & 0.0007 & 0.0007 & 94.50\% \\ \hline
\multirow{5}{*}{ML + GLMs - most data adaptive} & G-comp. & -0.0602 & 0.0010 & 0.0046 & - \\ \cline{2-6} 
 & IPTW & 0.1246 & 0.0095 & 0.0251 & 29.60\% \\ \cline{2-6} 
 & IPTW-DR & 0.1106 & 0.0080 & 0.0202 & 25.60\% \\ \cline{2-6} 
 & TMLE & 0.1184 & 0.0106 & 0.0246 & 24.70\% \\ \cline{2-6} 
 & CV-TMLE & 0.0008 & 0.0006 & 0.0006 & 94.10\% \\ \hline
\end{tabular}
\caption{Performance metrics of each estimator $\hat{\psi}_{d^*_n}$ (G-computation, IPTW, IPTW-DR, TMLE) for $\psi_{0,d^*_n}$ or $\hat{\psi}_{d^*_{n,v}}$ (CV-TMLE) for  $\psi_{0,d^*_{n,v}}$, for each library configuration of $Q_n$ and $d^*_n$.}
\end{table}

\begin{figure}[h]
    \centering
    \includegraphics[scale = .35]{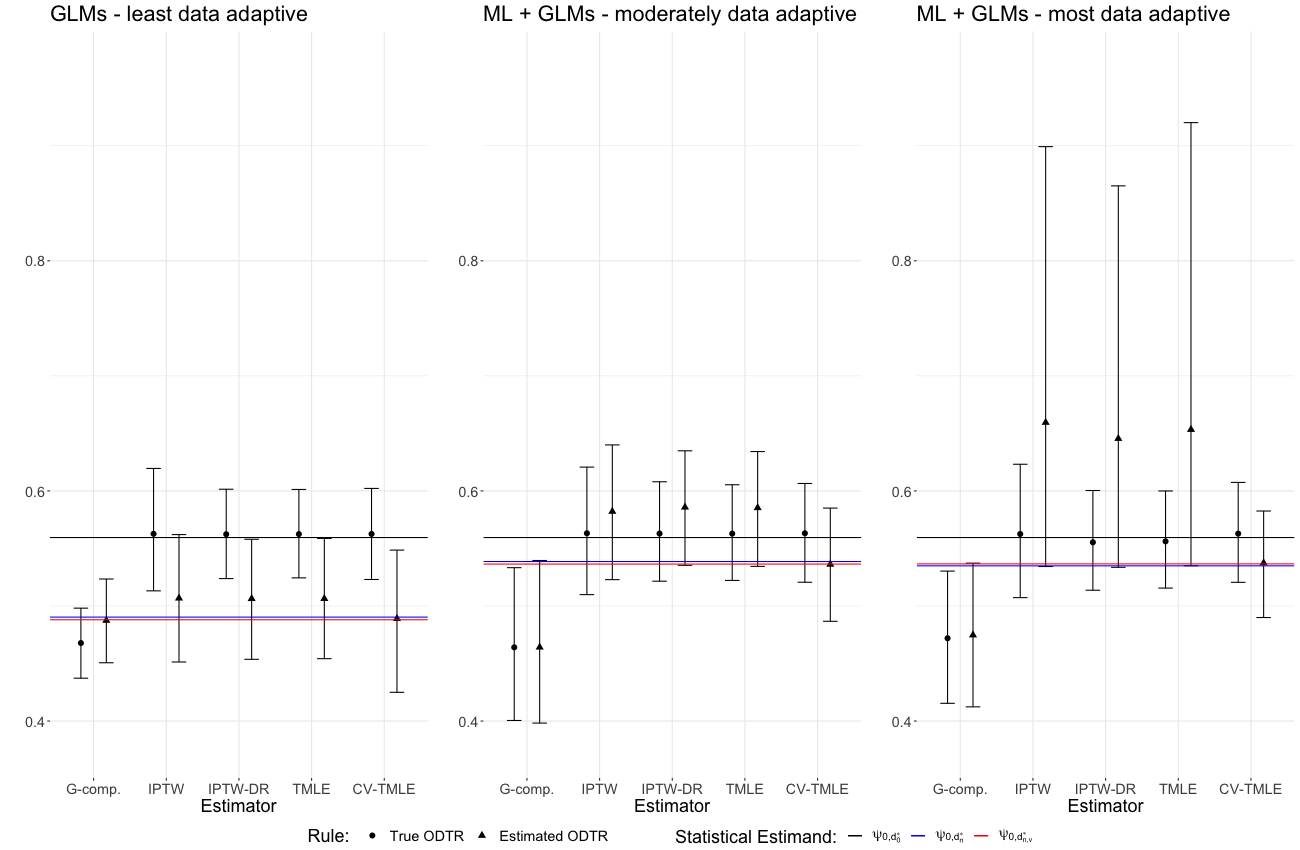}
    \caption{Performance of candidate estimators of the value of a given rule, evaluated for 3 SuperLearner library configurations with increasing (left to right) levels of data-adaptivity used for estimating the true outcome regression $Q_0$ and/or the true optimal rule $d^*_0$ (``GLM - least data adaptive", ``ML + GLMs - moderately data adaptive", "ML + GLMs - most data adaptive"). The horizontal black line depicts the true mean outcome under the true ODTR $\psi_{0,d^*_0}$; the blue and red lines are true values of the data-adaptive parameters $\psi_{0,d^*_n}$ and $\psi_{0,d^*_{n,v}}$, respectively, averaged over each of the 1,000 simulated samples. Points with error bars show the distribution of the estimators across the 1,000 simulated samples (G-computation estimator, IPTW estimator, TMLE, and CV-TMLE); the points (circles and triangles) show the estimates averaged over the samples, and error bars show the $2.5^{th}$ and $97.5^{th}$ quantiles of the distribution of each estimator across the simulation repetitions. The circles depict the estimators under a known rule $\hat{\psi}_{d=d_0^*}$ and the triangles illustrate the estimators under an estimated rule, either $\hat{\psi}_{d^*_n}$ or $\hat{\psi}_{d_{n,v}^*}$ (for CV-TMLE). DGP used here includes covariates that are dependent.}
    \label{plotfig2}
\end{figure}

\end{document}